\documentclass[12pt]{article} 

\usepackage{graphics,color}
\usepackage{amssymb}
\usepackage{amsbsy}
\usepackage{epsfig}

\newcommand{\be}{\begin{equation}}
\newcommand{\ee}{\end{equation}}
\newcommand{\bea}{\begin{eqnarray}}
\newcommand{\eea}{\end{eqnarray}}
\newcommand{\bit}{\begin{itemize}}
\newcommand{\eit}{\end{itemize}}
\newcommand{\bc}{\begin{center}}
\newcommand{\ec}{\end{center}}
\newcommand{\bra}{\langle}
\newcommand{\ket}{\rangle}

\newcommand{\si}{\,\mbox{$\sum$}\hs{-0.47cm}\int}
\newcommand{\sgn}{\mbox{sgn}}
\newcommand{\im}{{\mathrm{Im}}}
\newcommand{\re}{{\mathrm{Re}}}
\newcommand{\Tr}{{\mathrm{Tr}}}

\newcommand{\cO}{{\cal O}}

\newcommand{\cD}{{\cal D}}

\newcommand{\C}{{\cal C}}
\newcommand{\om}{\omega}

\newcommand{\tw}{\Gamma}

\newcommand{\po}{p^{0}}
\newcommand{\qo}{q^{0}}

\newcommand{\lv}{{\mathbf l}}
\newcommand{\pv}{{\mathbf p}}
\newcommand{\kv}{{\mathbf k}}

\newcommand{\rv}{{\mathbf r}}

\newcommand{\pvuni}{\hat{{\mathbf p}}}

\newcommand{\rvuni}{\hat{{\mathbf r}}}

\newcommand{\hm}{\hspace*{-0.6cm}}
\newcommand{\hs}[1]{\hspace*{#1}}

\newcommand{\simg}{\gtrsim}
\newcommand{\siml}{\lesssim}

\newcommand{\half}{\frac{1}{2}}

\newcommand{\ep}{\epsilon}

\newcommand{\bean}{\begin{eqnarray*}}
\newcommand{\eean}{\end{eqnarray*}}
\newcommand{\nn}{\nonumber}
\newcommand{\veck}{{\mathbf k}}
\newcommand{\vecp}{{\mathbf p}}

\newcommand{\vecx}{{\mathbf x}}
\newcommand{\vecy}{{\mathbf y}}

\newcommand{\vecnul}{{\mathbf 0}}
\begin{document}

\title{
\vskip -100pt
{
\begin{normalsize}
\mbox{} \hfill  hep-ph/0402192
\vskip  70pt
\end{normalsize}
}
{\bf\Large
Shear Viscosity in the $O(N)$ Model
}}

\author{
Gert Aarts\thanks{email: 
aarts@mps.ohio-state.edu}\addtocounter{footnote}{1}
 {} and
J.\ M.\ Mart{\'\i}nez Resco\thanks{email: marej@mps.ohio-state.edu}
 \\ {} \\
 {\em\normalsize  Department of Physics, The Ohio State University} \\
 {\em\normalsize  174 West 18th Avenue, Columbus, OH 43210, USA}
}

\date{February 18, 2004}
\maketitle

\begin{abstract}
 We compute the shear viscosity in the $O(N)$ model at first nontrivial
order in the large $N$ expansion. The calculation is organized using the
$1/N$ expansion of the 2PI effective action (2PI-$1/N$ expansion) to
next-to-leading order, which leads to an integral equation summing ladder
and bubble diagrams.  We also consider the weakly coupled theory for
arbitrary $N$, using the three-loop expansion of the 2PI effective action.  
In the limit of weak coupling and vanishing mass, we find an approximate
analytical solution of the integral equation.  For general coupling and
mass, the integral equation is solved numerically using a variational
approach.  The shear viscosity turns out to be close to the result
obtained in the weak-coupling analysis.

\end{abstract}

\newpage



\section{Introduction}

Due to the recent progress in relativistic heavy ion collisions and
cosmology, detailed theoretical investigations of the dynamics of quantum
fields out of equilibrium have become an active subject of research. A
successful approach to solve the dynamics of quantum fields far from
equilibrium as well as the subsequent stage of equilibration and
thermalization makes use of the so-called two-particle-irreducible (2PI)
effective action (see Refs.\ \cite{Berges:2000ur}-\cite{Ikeda:2004in} for
recent nonequilibrium applications).

The final stages of thermalization in systems out of equilibrium with
conserved quantities can be described by hydrodynamics, characterized by
an equation of state and transport coefficients. Recently, we investigated
the connection between the 2PI effective action and transport coefficients
for a variety of field theories and found that the lowest nontrivial
truncation of the 2PI effective action determines correctly transport
coefficients in a weak coupling or $1/N$ expansion at leading
(logarithmic) order \cite{Aarts:2003bk}.\footnote{For the bulk viscosity
it is necessary to go to higher-order truncations \cite{Calzetta:1999ps}.}
We emphasized that this result provides an important benchmark to validate
commonly used truncation schemes for nonequilibrium quantum field
dynamics.\footnote{In this context it would be interesting to compute
transport coefficients within the so-called 2PPI effective action approach
\cite{Baacke:2002ee}.}

In applications to heavy ion dynamics, successful hydrodynamical
descriptions of heavy ion collisions are so far based on ideal
hydrodynamics, assuming infinitely fast thermalization and vanishing
transport coefficients \cite{Kolb:2003dz}. The extension to nonideal
hydrodynamics is nontrivial, but the effects of viscous corrections are
currently under investigation \cite{Teaney:2003pb}. In
order to make further progress, it is crucial to know the magnitude of
transport coefficients quantitatively \cite{Teaney:2003pb}.

Transport coefficients in relativistic plasmas at high temperature can be
computed following different approaches. The first complete calculations
in hot gauge theories were done using kinetic theory: to leading
logarithmic order in the weak coupling expansion \cite{Arnold:2000dr}, to
full leading order \cite{Arnold:2002zm,Arnold:2003zc}, and in the large
$N_f$ limit \cite{Moore:2001fg}. Using field theory techniques, the shear
and bulk viscosities were obtained in a single-component scalar field
theory with cubic and quartic interactions through the summation of an
infinite series of ladder diagrams \cite{Jeon:if} (for a more concise
analysis of the shear viscosity, see \cite{Wang:2002nb}). In
Ref.~\cite{ValleBasagoiti:2002ir} a simple and economical way was
presented to carry out the sum of ladder diagrams contributing to the
shear viscosity to leading order in a quartic scalar theory and to the
shear viscosity and electrical conductivity in the leading-log
approximation in (non)abelian gauge theories. The Ward identity in the
calculation of the electrical conductivity in QED was studied in
Ref.~\cite{Aarts:2002tn}, while an alternative diagrammatic approach
employing a dynamical renormalization group to study the conductivity was
presented in Ref.~\cite{Boyanovsky:2002te}.  The prospects of extracting
transport coefficients nonperturbatively using lattice QCD have been
discussed in Ref.~\cite{Aarts:2002cc} and first results have been
obtained~\cite{Gupta:2003zh} (see also \cite{Karsch:1986cq}). In strongly
coupled supersymmetric Yang-Mills theories, the shear viscosity has been
computed with the help of the AdS/CFT
correspondence~\cite{Policastro:2001yc}. Finally, the shear viscosity has
been computed in the hadronic phase using phenomenological
models~\cite{Muronga:2003tb,Dobado:2003wr}.

\begin{figure}[t]
 \begin{center}
 \epsfig{figure=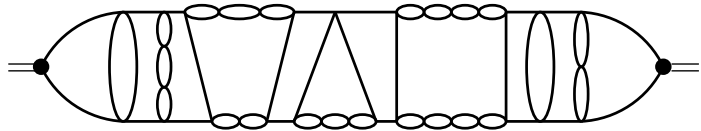,height=1.7cm}
 \end{center}
 \vspace{-0.5cm}
 \caption{Typical skeleton diagram that contributes to the shear viscosity
 in the $O(N)$ model at first nontrivial order in the large $N$ limit.
 The black dots with tiny lines represent the external bilinear operators.
}
 \label{figaladder}
\end{figure}

In this paper we consider the $O(N)$ model in its symmetric phase and
compute the shear viscosity to first nontrivial order in the $1/N$
expansion. The $O(N)$ model is widely used when applying methods and
techniques in thermal and nonequilibrium field theory, because without 
running into the issue of gauge invariance, it still leaves a nontrivial 
problem
to solve. Moreover, it acts as a low energy effective description of QCD
(for $N=4$) and is frequently used in early cosmology (inflation,
reheating).  The large $N$ expansion offers the possibility to explore the
behavior of transport coefficients outside the weak-coupling domain and in
combination with the 2PI effective action, it naturally leads to the
inclusion of the required medium effects in the single particle
propagators. We use a field theory approach to compute the shear
viscosity. As we will show below, a typical ladder diagram that
contributes at leading order is shown in Fig.~\ref{figaladder}. We note
that the lines in this diagram are dressed propagators, so it should be
regarded as a skeleton diagram. As we have recently shown
\cite{Aarts:2003bk}, the summation of these classes of diagrams is neatly
organized using the 2PI effective action. In particular, the 2PI-$1/N$
expansion to next-to-leading order (NLO) \cite{Berges:2001fi,Aarts:2002dj}
sums precisely the relevant set of diagrams.

The paper is organized as follows. In Section~\ref{2pi} we describe the
2PI effective action for the $O(N)$ model and write down the ensuing
integral equation at next-to-leading order in the $1/N$ expansion. We
argue that it sums all necessary diagrams that contribute to the shear
viscosity at leading order. In the next Section we study the various
elements that appear in the integral equation and which are essential for
its calculation: the single particle spectral function, the gap equation
for the mass and its renormalization, the auxiliary correlator summing the
chain of bubbles diagrams, and the thermal width. In
Section~\ref{eta} we derive a compact expression for the shear viscosity
in terms of an effective vertex summing the ladder diagrams.  
Section~\ref{inte} is devoted to the integral equation for the effective
vertex. We show how to cast the problem of solving it into a variational
one, better suited for numerical analysis. In Section~6 we study the
weak-coupling limit of the integral equation and find an analytical result
in the limit of ultrahard momentum. In Section~\ref{num} we solve
numerically the variational problem and present the results for the shear
viscosity. The final Section is devoted to the conclusions.  In Appendix A
we consider the weakly coupled $O(N)$ model for arbitrary $N$ using the
three-loop expansion of the 2PI effective action.


\section{2PI-$1/N$ expansion} 
\label{2pi}

We consider a real scalar $N$--component quantum field
$\phi_a$ ($a=1,\ldots, N$) with a classical $O(N)$--invariant action,
\be
 S[\phi] = \int_x \left[ 
 \frac{1}{2} \partial_{0} \phi_a\partial_{0} \phi_a
 -\frac{1}{2} \partial_{i} \phi_a\partial_{i} \phi_a 
 -\frac{1}{2} m_{0}^2 \phi_a \phi_a
 -\frac{\lambda_0}{4! N} \left(\phi_a\phi_a\right)^2 \right].
\ee
The mass parameter $m_0$ and coupling constant $\lambda_0$ are bare 
parameters. To the order we are working, field renormalization is not 
necessary. We use the notation
\be
\int_x = \int_\C dx^0 \int d^3 x,
\ee
where $\C$ is a contour in the complex-time plane. We will work in the
imaginary time formalism, so below we specialize to the Matsubara contour,
running from 0 to $-i/T$.

According to the Kubo formula, the shear viscosity can be obtained from 
the slope of a spectral function at zero frequency, 
\be
 \label{eqetadef}
 \eta = \frac{1}{20}\frac{\partial}{\partial q^0}
 \rho_{\pi\pi}(q^0,\vecnul)\Big|_{q^0=0}, 
\ee 
where 
\be 
 \rho_{\pi\pi}(x-y)=\bra [\pi_{ij}(x), \pi_{ij}(y)]\ket, 
\ee 
with $\pi_{ij}$ the traceless part of the spatial energy-momentum tensor, 
\be \label{pi} 
 \pi_{ij}(x)=\partial_i\phi_a(x)\partial_j\phi_a(x) - \frac{1}{3}
 \delta_{ij}\partial_k\phi_a(x)\partial_k\phi_a(x). 
\ee 

It is well known from weak coupling studies that a one loop calculation of
the shear viscosity is incorrect, since diagrams that appear at higher
order in the loop expansion contribute at leading order and the
computation must be carried out using dressed
propagators~\cite{Jeon:if,ValleBasagoiti:2002ir}. For a single-component
scalar field ($N=1$), Jeon showed that these higher order diagrams are
ladder diagrams, where the rung in the ladder is a single
bubble~\cite{Jeon:if}. Due to the kinematical configuration, the
propagators on the side rails suffer from pinching poles. As a result the
thermal width, which is determined by the imaginary part of the self
energy, has to be included in these propagators, while the real part is
subleading and can be neglected (at weak coupling).  The picture is quite
similar for gauge theories at leading logarithmic order in the weak
coupling limit \cite{ValleBasagoiti:2002ir,Aarts:2002tn}. The ladder
series is conveniently summed through an integral equation for an
effective vertex~\cite{Jeon:if,Wang:2002nb,ValleBasagoiti:2002ir}. We have
shown recently that this integral equation in the required kinematic
configuration appears naturally in the 2PI effective action
formalism~\cite{Aarts:2003bk}. In the $1/N$ expansion, in which we are
interested here, the presence of pinching poles also leads to higher order
diagrams in the loop expansion contributing at leading order, and the 2PI
effective action formalism neatly organizes this calculation as well, as
we describe now.

The 2PI effective action is a functional of the time-ordered two-point 
function 
$
G_{ab}(x,y)  =  \langle T_{\cal C}\phi_a(x)\phi_b(y)\rangle
$
and can be parametrized as \cite{Cornwall:1974vz,Luttinger}
\be
\Gamma[G] = \frac{i}{2}\Tr \ln G^{-1} + \frac{i}{2}\Tr\, G_0^{-1}(G-G_0)
+\Gamma_2[G].
\ee
Throughout we consider $\bra \phi_a(x)\ket = 0$. The classical inverse 
propagator $i G_{0}^{-1}$ is given by
\be
i G^{-1}_{0,ab}(x,y)
= - \left[ \square_x + m_0^2 \right] \delta_{ab} \delta_{\C}(x-y),
\ee
with $\delta_{\C}(x-y) \equiv\delta_{\C}(x^0-y^0) \delta(\vecx- \vecy)$.
$\Gamma_2[G]$ is the sum of all 2-particle irreducible (2PI) diagrams with 
no external legs and exact propagators on the internal lines.  
Extremizing this effective action,
\be
\frac{\delta \Gamma[G]}{\delta G_{ab}(x,y)} = 0,
\ee
leads to a Dyson equation for the two-point function
\be
\label{eqgap}
G_{ab}^{-1}(x,y) = G_{0,ab}^{-1}(x,y) - \Sigma_{ab}(x,y)
\ee
where the self energy
\be
\label{eqself}
\Sigma_{ab}(x,y) \equiv
2i\frac{\delta \Gamma_2[G]}{\delta G_{ab}(x,y)}
\ee
depends on the full propagator $G$.

\begin{figure}[t]
 \begin{center}
 \epsfig{file=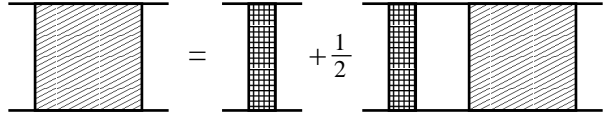,height=1.5cm}
 \end{center}
 \vspace{-0.5cm}
 \caption{Integral equation for the 4-point function.}
 \label{figintegral}
\end{figure}

Within the 2PI formalism, a 4-point vertex function $\Gamma^{(4)}$
appears, defined as the usual connected 4-point function with the 
external legs removed. This 4-point function obeys an integral equation 
that can be obtained using standard functional relations \cite{Cornwall:1974vz} 
(see Fig.\ \ref{figintegral})
\bea \nn
 &&\hm \Gamma^{(4)}_{ab;cd}(x,y;x',y') = \Lambda_{ab;cd}(x,y;x',y') \\
 \label{eqfour}
 &&\hm + \half \int_{ww'zz'}
 \Lambda_{ab;ef}(x,y;w,z) G_{ee'}(w,w') G_{ff'}(z,z') 
 \Gamma^{(4)}_{e'f';cd}(w',z';x',y').
\eea
The kernel or rung $\Lambda$ is defined as
\be \label{eqlambda}
 \Lambda_{ab;cd}(x,y;x',y') = 2 \frac{\delta \Sigma_{ab}(x,y)}{\delta 
 G_{cd}(x',y')}.
\ee
Semicolons separate indices with a different origin \cite{Aarts:2003bk}.

\begin{figure}[b]
 \begin{center}
 \epsfig{figure=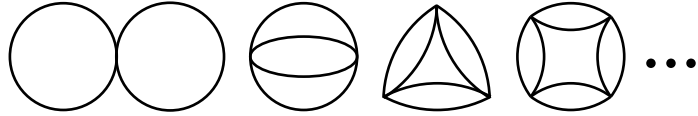,height=1.5cm}
 \end{center}
 \vspace{-0.5cm}
 \caption{Contributions to the 2PI effective action in the $O(N)$ model in
 the 2PI--$1/N$ expansion at LO and NLO. Only the first few diagrams at 
 NLO are shown.}
 \label{figN}
\end{figure}

\begin{figure}[t]
 \begin{center}
 \epsfig{figure=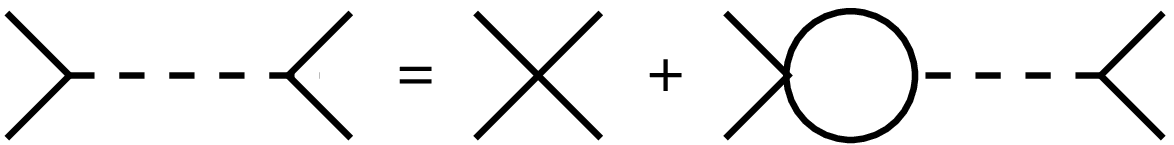,height=0.8cm}
 \end{center}
 \vspace{-0.5cm}
 \caption{Integral equation for the auxiliary correlator $D$.}
 \label{figintegralD}
\end{figure}

\begin{figure}[t]
 \begin{center}
 \epsfig{figure=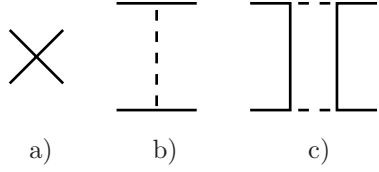,height=1.6cm} \\
  {\footnotesize \hspace*{-0.5cm} a) \hspace*{1.1cm} b) \hspace*{1.5cm} c) }
 \end{center}
 \vspace{-0.5cm}
 \caption{4-point kernel in the 2PI--$1/N$ expansion of the $O(N)$ model 
at LO and NLO.}
 \label{figkernelN}
\end{figure}

The relations so far are general. We now specialize to the 2PI-$1/N$ 
expansion to NLO. The 2PI part of the effective action can be written as 
\cite{Berges:2001fi,Aarts:2002dj}
\be
\Gamma_2[G] = \Gamma_2^{\rm LO}[G] + \Gamma_2^{\rm NLO}[G] + \ldots,
\ee
where (see Fig.\ \ref{figN})
\bea \label{eqGammaLO}
 \Gamma_2^{\rm LO}[G] =&&\hm -\frac{\lambda_0}{4!N}\int_x G_{aa}(x,x)
 G_{bb}(x,x),\\
 \label{eqGammaNLO}
 \Gamma_2^{\rm NLO}[G] =&&\hm \frac{i}{2}\Tr \ln {\rm\mathbf{B}},
\eea
with 
\be
{\rm\mathbf{B}}(x,y) = \delta_\C(x-y) - \frac{i\lambda_0}{3N}\Pi(x,y).
\ee
The function {\bf B} depends on the single bubble diagram defined by 
\be
\Pi(x,y) = -\frac{1}{2}G_{ab}(x,y)G_{ab}(x,y).
\ee
Expanding the logarithm in Eq.\ (\ref{eqGammaNLO}) generates the closed 
chain of bubble diagrams as in Fig.\ \ref{figN}.
The self energy follows from Eq.~(\ref{eqself}). We write  
$\Sigma = \Sigma^{\rm LO} + \Sigma^{\rm NLO}$, with 
\begin{figure}
 \centerline{\epsfig{figure=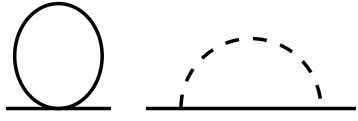,height=1.5cm}}
 \caption{Contributions to the self energy at LO and NLO. The auxiliary 
correlator $D$ is represented by the dashed line.}
\label{figselfenergyN}
\end{figure}
\bea
\Sigma^{\rm LO}_{ab}(x,y) =&&\hm -i\frac{\lambda_0}{6N}\delta_{ab}
G_{cc}(x,x) \delta_\C(x-y), \\
\Sigma^{\rm NLO}_{ab}(x,y) =&&\hm -G_{ab}(x,y)D(x,y),
\eea
shown in Fig.\ \ref{figselfenergyN}. Here we introduced the auxiliary 
correlator 
\be
D(x,y) = \frac{i\lambda_0}{3N} {\rm\mathbf{B}}^{-1}(x,y).
\ee
{}From the identity {\bf{B}}$^{-1}${\bf{B}} $=1$ it follows that 
the auxiliary correlator obeys
\be
D(x,y) = \frac{i\lambda_0}{3N} \left[ \delta_\C(x-y) + \int_z 
\Pi(x,z)D(z,y)
\right],
\ee
which is depicted in Fig.\ \ref{figintegralD}. 
For the rung, we write $\Lambda=\Lambda^{\rm LO}+\Lambda^{\rm NLO}$, 
and find (see Fig.~\ref{figkernelN})
\bea
 \Lambda^{\rm LO}_{ab;cd}(x,y; x',y') =&&\hm -\frac{i\lambda_0}{3N}
 \delta_{ab}\delta_{cd}\delta_\C(x-y)\delta_\C(x-y')\delta_\C(x'-y), \\
 \nn\Lambda^{\rm NLO}_{ab;cd}(x,y;x',y') =&&\hm
 -\left[\delta_{ac}\delta_{bd} + \delta_{ad}\delta_{bc}\right]
 D(x,y) \delta_\C(x-x')\delta_\C(y-y') \\
 && + 2 G_{ab}(x,y) D(x,x') D(y,y') G_{cd}(x',y').
\eea

Finally, we specialize to the $O(N)$--symmetric case, take $G_{ab}(x,y) 
= \delta_{ab}G(x,y)$ and $\Sigma_{ab}(x,y) = \delta_{ab}\Sigma(x,y)$, choose 
the Matsubara contour and write the equations in momentum space.
The final set of equations then reads
\be  \label{eqGeucl}
 G(P) = \frac{1}{ \om_n^2+\pv^2 + m_0^2 +\Sigma(P)},
\ee
with
\bea \label{eqSigmaLO}
 \Sigma^{\rm LO}(P) =&&\hm \frac{\lambda_0}{6}\si_K G(K), \\
     \label{eqSigmaNLO}
 \Sigma^{\rm NLO}(P) =&&\hm -\si_K  G(P+K)D(K), 
\eea
and
\bea \label{eqDeucl}
 D(P) =&&\hm \frac{1}{-3N/\lambda_0 + \Pi(P)},\\
     \label{eqPi}
 \Pi(P) =&&\hm -\frac{N}{2} \si_K G(P+K)G(K).
\eea
The 4-point function obeys
\be  \label{intequ}
 \Gamma^{(4)}_{ab;cd}(R,K) = \Lambda_{ab;cd}(R,K) + \half \si_{P} 
 \Lambda_{ab;ef}(R,P) G^2(P) \Gamma^{(4)}_{ef;cd}(P,K),
\ee
with the kernel
\bea \label{eqLambdaLO}
 \Lambda^{\rm LO}_{ab;cd}(R,P) =&&\hm -\frac{\lambda_0}{3N}
 \delta_{ab}\delta_{cd}, \\
 \nn
 \Lambda^{\rm NLO}_{ab;cd}(R,P) =&&\hm
 \left[\delta_{ac}\delta_{bd} + \delta_{ad}\delta_{bc}\right] D(R-P) 
 \\ &&\hm 
    \label{eqLambdaNLO}
 + 2 \delta_{ab}\delta_{cd} \si_L  G(R-L) D^2(L) G(L-P).
\eea
Here we used the notation $P=(i\om_n, \vecp)$, where $\om_n = 2\pi nT$ 
($n\in\mathbb{Z}$) is the Matsubara frequency, and
\be
 \si_K  = T\sum_n \int_\veck, \;\;\;\;\;\;\;\;\;\;\;\; 
 \int_\veck = \int \frac{d^3k}{(2\pi)^3}.
\ee

\begin{figure}[t]
 \begin{center}
 \epsfig{figure=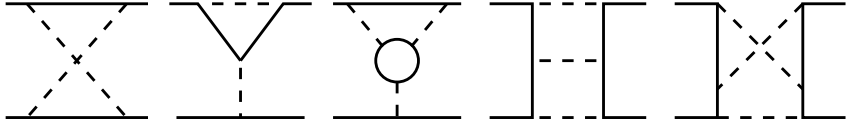,height=1.6cm}
 \end{center}
 \vspace{-0.5cm}
 \caption{Examples of rungs that may contribute at 
 next-to-leading order to the shear viscosity in the $1/N$ expansion.}
 \label{figkernelNNLO} 
\end{figure}

In the remainder of this section, we argue that the set of Eqs.\
(\ref{eqGeucl})-(\ref{eqLambdaNLO}) sums all the diagrams contributing to
the shear viscosity at leading order in the $1/N$ expansion. First we note
that the one loop contribution to the viscosity is proportional to
$N^{2}$. Here, one factor of $N$ arises from the group indices running in
the loop, the other originates from the pair of pinching poles in the
loop. Pinching poles are screened by the imaginary part of the retarded
self energy $\Sigma$, which appears first at NLO (the leading order 
contribution to $\Sigma$ is 
real). Since pinching poles are sensitive to the inverse of the imaginary 
part, each pair of pinching poles gives a contribution that scales as $N$.
Therefore the shear viscosity takes the form\footnote{Note that the ratio 
of the shear viscosity and the entropy density $s$ is proportional to $N$, 
and therefore always far above the lower bound conjectured recently, 
$\eta/s\geq 1/4\pi$ \cite{Kovtun:2003wp}.}
\be  \label{eqeta}
 \eta=N^{2}T^{3}\left[F\left(\lambda(\mu), \frac{m_{R}}{T},
 \frac{\mu}{T}\right) 
 +\cO\left(\frac{1}{N}\right)\right],
\ee
where $m_{R}$ is the renormalized mass
in vacuum and $\lambda(\mu)$ is the renormalized coupling constant at the
scale $\mu$. The shear viscosity is independent of $\mu$. It 
is now straightforward to identify which diagrams contribute to the shear
viscosity at leading order. The auxiliary correlator $D$ is proportional
to $1/N$, see Eqs.\ (\ref{eqDeucl}, \ref{eqPi}). Starting from the naive
one-loop expression, adding a vertical $D$ correlator as in Fig.\
\ref{figkernelN}b yields one extra pair of pinching poles that cancels the
explicit $1/N$ from the $D$ correlator. Therefore all vertical line
insertions contribute at the same order. An insertion of the box rung (see
Fig.\ \ref{figkernelN}c)  results in two additional $D$ correlators, one
additional pair of pinching poles, and one additional closed loop over the
group indices. Therefore also all box rung insertions contribute at the
same order. Finally, the lowest order rung (see Fig.\ \ref{figkernelN}a),
which would generate a string of bubbles, does not contribute to the shear
viscosity.  This was shown by Jeon \cite{Jeon:if} for the weakly coupled
single component case and will be confirmed below. Rungs that are down by
$1/N$ have additional $D$ correlators but no compensating additional
pinching poles or closed loops.  Some examples of subleading rungs are
shown in Fig.\ \ref{figkernelNNLO}. We find therefore that the typical
diagram contributing at leading order is as shown in
Fig.~\ref{figaladder}. Throughout the paper we neglect subleading powers
of $N$ as indicated in Eq.~(\ref{eqeta}).

Before analyzing the integral equation, which sums the required
contributions, several elements are needed in detail. We study those in
the next section.


\section{Quasiparticles}     \label{quasi}

In this section we study the single particle propagators that are dressed
to account for plasma effects, which are crucial to the calculation of
transport coefficients. A convenient way to study these effects is through
the use of the single particle spectral density.

Analytically continuing the euclidean propagator (\ref{eqGeucl})
to real frequencies yields the retarded and advanced propagators,
\bea
\nn
G_R(P) =&&\hm G(i\om_n\to p^0+i0,\vecp) = G_A^*(P) \\
 = &&\hm \frac{1}{-p_0^2+\pv^2+m_0^2+\re\Sigma_R(P)
+ i\im\Sigma_R(P)},
\eea
from which the spectral function is obtained as
\bea
\nn
\rho(P) =&&\hm  -i\left[ G_R(P) - G_A(P) \right] \\
= &&\hm  \frac{-2\im\Sigma_R(P)}{
\left[p_0^2-\pv^2-m_0^2-\re\Sigma_R(P)\right]^2 
+ \left[\im\Sigma_R(P)\right]^2}.
\eea
 The leading part of the self energy $\Sigma^{\rm LO}$ is of the same
order as the mass in the $1/N$ expansion and hence it cannot be neglected.  
The real part of $\Sigma^{\rm NLO}_R$ on the other hand can be dropped
since it is suppressed by $1/N$ compared to the leading
piece.\footnote{This is the reason field renormalization does not
enter in this problem.} Since $\im\Sigma_R$ is proportional to $1/N$, the
spectral density generically has a simple on-shell form in the large $N$ 
limit,
 \be   \label{eqrhofree}
 \rho(P) = 2\pi\sgn(p^0)\delta(p_0^2-\om_\pv^2) +\cO(1/N), 
\ee 
with 
\be
 \om_\pv = \sqrt{\pv^2 +M^2}, \;\;\;\;\;\;\;\; M^2 = m_0^2 + \Sigma^{\rm LO}.
\ee
However, when pairs of propagators with pinching poles are present, the
imaginary part cannot be neglected as it separates the poles in the
complex-energy plane that approach the real axis from below and
above \cite{Jeon:if,ValleBasagoiti:2002ir}. In
this case, the full propagator $G(P)$ has a cut
along the whole real $p^0$ axis since $\im \Sigma_R(P)$ is
nonvanishing for all real $p^0$. Moreover, the product of retarded
and advanced propagators is directly proportional to its inverse,
 \be   \label{eqPP1}
 G_{R}(P)G_{A}(P) = 
 \frac{1}{\left[p_{0}^{2}-\pv^{2}-m_{0}^{2}-\re\Sigma_{R}(P)\right]^{2} 
 + \left[\im\Sigma_{R}(P)\right]^{2}}
 = \frac{\rho(P)}{-2\im\Sigma_R(P)}. 
\ee
In the limit of large $N$ we may use Eq.\ (\ref{eqrhofree}) and this 
expression reduces to
\be  \label{eqPP}
 G_R(P) G_A(P) = \frac{\rho(P)}{2p^0\Gamma_\pv} +\cO(1),
\ee
with $\Gamma_\pv$ the (on-shell) thermal width defined from the imaginary 
part of the retarded self energy as
\be
 \Gamma_\pv = -\frac{\im \Sigma_R(P)}{p^0}\Big|_{p^0=\pm\om_\pv}.
\ee
 In the remainder of this section we study the corrections introduced by
the self energy. The real part of $\Sigma$ leads to a gap equation for
the mass which requires renormalization of both the mass parameter and
coupling constant. The imaginary part is computed in terms of the
auxiliary correlator $D$, which we work out in detail.


\subsection{Gap equation and renormalization}

Since the real part of $\Sigma_R^{\rm NLO}$ can be systematically 
neglected, the gap equation for the mass parameter $M$ reads
\be
 M^2 = m_0^2 + \frac{\lambda_0}{6}\si_P G(P),
\ee
with
\be
 G(P) = \frac{1}{ \om_n^2+\pv^2 + M^2}.
\ee
 This gap equation is divergent and in order to renormalize
it\footnote{See Refs.\ \cite{vanHees:2001ik,Blaizot:2003br} for recent
studies of renormalization in the 2PI effective action formalism.} we
also need the integral equation for the 4-point function,
Eq.~(\ref{intequ}), at lowest order. In this approximation the 4-point 
function is momentum independent and Eq.~(\ref{intequ}) can be solved as
 \be  \label{eqGamma4}
 \Gamma^{(4)}_{ab;cd}(R,K) = \delta_{ab}\delta_{cd}\Gamma^{(4)}, 
\hs{1.3cm} 
 \frac{1}{\Gamma^{(4)}} =  -\frac{3N}{\lambda_0} + \Pi(0),
\ee
with the single bubble at zero momentum
\be
 \Pi(0) = -\frac{N}{2} \si_{P}  G^2(P).
\ee
 We note that the equation for the 4-point function is identical to the
equation for the auxiliary correlator $D(P)$ at zero momentum.  Therefore
the renormalization carried out to obtain a finite gap equation renders
$D(P)$ finite too.

In order to regulate the divergent integrals we use dimensional 
regularization in $3-2\ep$ dimensions. Both $\lambda_0$ 
and $\Gamma^{(4)}$ have dimension $\mu^{2\ep}$. We introduce a 
dimensionless coupling $\lambda_R$ via
\be
 \Gamma^{(4)} = -\frac{\lambda_{R}\mu^{2\epsilon}}{3N},
\ee
such that the equations to renormalize read
\bea
 \frac{1}{\lambda_{R}} = &&\hm \frac{\mu^{2\epsilon}}{\lambda_0}
 +\frac{\mu^{2\epsilon}}{6}\si_P G^2(P), 
 \\ 
 M^2 =&&\hm  m_0^2 +\frac{\lambda_0}{6} \si_P G(P). \label{mass}
\eea
Since renormalization can be carried out at zero temperature, we only
compute the bubble in the vacuum
\be
 \frac{1}{\lambda_{R}}=\frac{\mu^{2\epsilon}}{\lambda_0}
 +\frac{1}{96\pi^{2}}\left(\frac{1}{\epsilon}
 +\ln(4\pi)-\gamma_{E}+2\ln\frac{\mu}{m_{R}}\right),
\ee
where $m_{R}$ is the renormalized mass at zero temperature.
Introducing the running coupling constant in the $\overline{MS}$ scheme  
$\lambda(\mu)$ via
\be \label{lrenor}
 \frac{1}{\lambda(\mu)}\equiv\frac{\mu^{2\epsilon}}{\lambda_0}
 +\frac{1}{96\pi^{2}}\left(\frac{1}{\epsilon}+\ln(4\pi)-\gamma_{E}\right),
\ee
we find 
\be
\frac{1}{\lambda_{R}} = 
\frac{1}{\lambda(\mu)}+\frac{1}{48\pi^{2}}\ln\frac{\mu}{m_{R}}.
\ee
The running coupling $\lambda(\mu)$ obeys the usual renormalization group 
(RG) equation with the correct $\beta$ function for the $O(N)$ model in 
the large $N$ limit, $\beta(\lambda)=\lambda^{2}/(48\pi^{2})$.

\begin{figure}[t]
 \begin{center}
 \epsfig{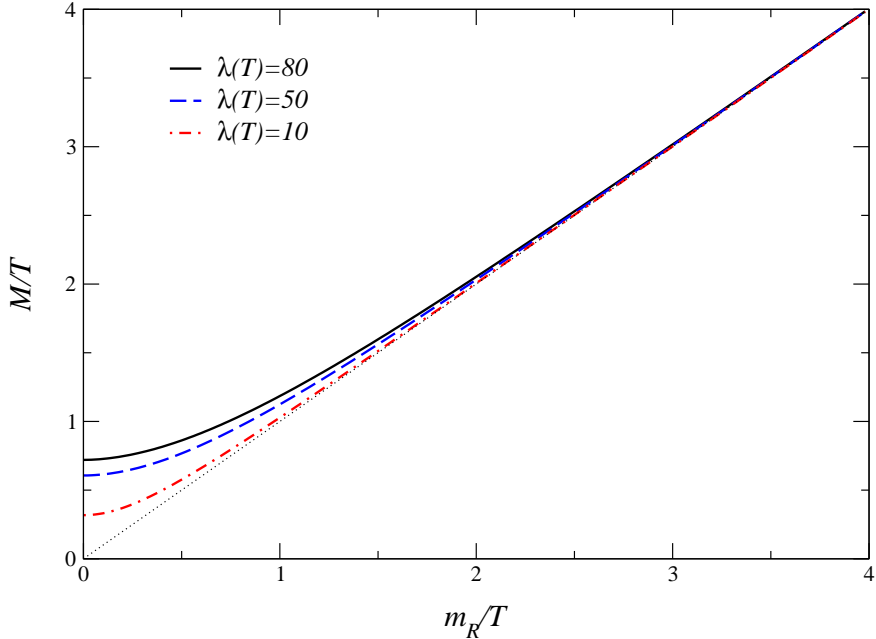}
 \end{center}
 \vspace{-0.5cm}
 \caption{Finite temperature mass $M/T$ as a function of the renormalized 
 mass at zero temperature $m_R/T$ for 3 different values of $\lambda(T)$.}
 \label{figmassmR}
\end{figure}

Returning to the gap equation (\ref{mass}) at zero temperature, we find 
\be \label{mrenor}
 m_{R}^{2} = m_{0}^{2} -
\frac{\lambda_{0}\mu^{-2\epsilon}}{6}\frac{m_{R}^{2}}{16\pi^{2}} 
 \left[
 \frac{1}{\epsilon} + 
  \ln(4\pi) -\gamma_{E} +1 + 2\ln \frac{\mu}{m_{R}} \right],
\ee
Combining Eqs.~(\ref{mass}, \ref{lrenor}, \ref{mrenor}), we arrive at the 
renormalized gap equation at finite temperature
\be
 M^{2}=m_{R}^{2}-\frac{\lambda(\mu)}{96\pi^{2}}
 \left[(M^{2}-m_{R}^{2})\left(1+\ln\frac{\mu^{2}}{m_{R}^{2}}\right) 
 +M^{2}\ln\frac{m_{R}^{2}}{M^{2}} \right] 
 +\frac{\lambda(\mu)}{6}\int_{\pv}\frac{n(\om_{\pv})}{\om_{\pv}},
\ee
where $n(\om)= 1/(e^{\om/T}-1)$ is the Bose distribution function.
The solution of this gap equation is independent of the renormalization 
scale $\mu$. A numerical solution of the gap equation for three values of 
$\lambda(\mu=T)$ is shown in Fig.\ \ref{figmassmR}.

There is one additional issue related to renormalization that needs to be
considered. This scalar theory has a Landau pole at the scale
$\Lambda_{L}=\mu e^{48\pi^{2}/\lambda(\mu)}$, where the running coupling
constant diverges. One has to require that all physical energy scales involved in
the problem, i.e.\ $m_R, M$, and $T$, are much smaller that the one
associated with the Landau pole.  This imposes a restriction on which
values of $\lambda(\mu)$ and $m_{R}$ can be considered. Taking, for example, $T$
to be the largest scale and demanding that $\Lambda_{L}/T\simg 40$, we
find that $\lambda(\mu=T)\siml 48\pi^{2}/\ln 40 \approx 128$.


\subsection{Auxiliary correlator}

Before proceeding with a calculation of the thermal width, we need to 
consider the chain of bubbles summed by the auxiliary correlator $D$.
The single bubble is given by
\be
 \Pi(P)=-\frac{N}{2}\si_{K}G(P+K)G(K),
\ee
which after performing the Matsubara sum reads
\bea 
\nn
 \Pi(P) =  -\frac{N}{2}\int_{\kv}  \frac{1}{4\om_{\kv}\om_{\rv}} 
 \left\{ 
\left[ n(\om_{\rv})-n(\om_{\kv}) \right]
 \left(\frac{1}{i\om_{n}+\om_{\kv}-\om_{\rv}}-
 \frac{1}{i\om_{n}-\om_{\kv}+\om_{\rv}}\right)
\right.  &&\hm
 \\
\label{bubble}
 \left. + 
\left[ 1+n(\om_{\kv})+n(\om_{\rv}) \right]
 \left( \frac{1}{i\om_{n}+\om_{\kv}+\om_{\rv}}-
 \frac{1}{i\om_{n}-\om_{\kv}-\om_{\rv}}\right)
\right\},  &&\hm
\eea
where $\rv=\kv+\pv$. In order to separate the logarithmic divergence in 
the zero-temperature contribution we write
\be
 \Pi(P)=\Pi_{0}(P)+\Pi_{T}(P), 
\;\;\;\;\;\;\;\; \Pi_{0}(P)=\Pi_{0}(0)+\Pi'_{0}(P),
\ee
where $\Pi_{0}(P)$ is the part without distribution functions.
The divergent contribution $\Pi_0(0)$ is similar to the one computed for 
the renormalization of the coupling constant,
\be
 \Pi_{0}(0)=-\frac{N}{32\pi^{2}}\left(\frac{1}{\epsilon}
 +\ln4\pi-\gamma_{E}+2\ln\frac{\mu}{M}\right),
\ee
while the remainder is finite and can be evaluated as
\be
 \Pi_0'(P) = -\frac{N}{16\pi^2}\left( 1+ \frac{1}{2}\sqrt{1+4M^2/P^2}
 \ln\frac{\sqrt{1+4M^2/P^2}-1}{\sqrt{1+4M^2/P^2}+1}\right),
\ee
with $P^2=\om_n^2+\pv^2$. The easiest way to arrive at the above result is 
to go back to the original four-dimensional euclidean integral. 

We are interested in the retarded bubble, obtained by analytical
continuation $i\om_n\to p^0+i0$.
For the finite contribution to the real part we write
\be
\re\Pi_R^f(P) = \re\Pi_0^{'R}(P) + \re\Pi_T^{R}(P),
\ee
with
\bea  \nn
 \re\Pi_0^{'R}(P) =&&\hm -\frac{N}{16\pi^2} \Bigg\{ 1 + \frac{1}{2}  
 \left[\Theta(s-4M^2)+ 
 \Theta(-s)\right] \beta(P)\ln\left|\frac{1-\beta(P)}{1+\beta(P)}\right|
 \\ &&\hm 
 - \Theta(4M^2-s)\Theta(s)B(P)\arctan\frac{1}{B(P)} \Bigg\},
\eea
where $s=p_0^2-p^2$, and  
\be
 \beta(P) = \sqrt{1-\frac{4M^2}{p_0^2-p^2}}, 
 \;\;\;\;\;\;\;\;\;\;\;\;
 B(P) = \sqrt{\frac{4M^2}{p_0^2-p^2}-1},
\ee
and
\be
 \re \Pi_T^R(P) = -\frac{N}{16\pi^2 p}\int_0^\infty dk\,
 \frac{k}{\om_\kv}n(\om_\kv)\ln\left|
 \frac{(k+p_+)(k+p_-)}{(k-p_+)(k-p_-)}\right|, 
\ee
with $p_\pm=\half\left[p\pm p^0\beta(P)\right]$.
The remaining integral can be done numerically.  

The imaginary part can be written in terms of single particle spectral 
functions as 
\be
\label{eqimpi1}
\im \Pi_R(P) =
-\frac{N}{4}\int_{K}
\rho(P+K)\rho(K)\left[ n(k^0) - n(p^0+k^0) \right],
\ee
where
\be
\int_{K} = \int \frac{d^4k}{(2\pi)^4},
\ee
and can be evaluated completely
\bea
\nn
\im\Pi_R(P) = &&\hm
- \Theta(s-4M^2) \frac{N}{32\pi} \left[ \beta(P) + \frac{2T}{p}
\ln \frac{1-e^{-\bar p_+/T}}{1-e^{-\bar p_-/T}} \right] \\
\label{eqimPi}
&&\hm
- \Theta(-s) \frac{NT}{16\pi p} \ln \frac{1-e^{-\bar
p_+/T}}{1-e^{\bar p_-/T}},
\eea
with $\bar p_\pm = \half\left[p^0\pm p\beta(P)\right]$.

\begin{figure}[t]
 \begin{center}
 \epsfig{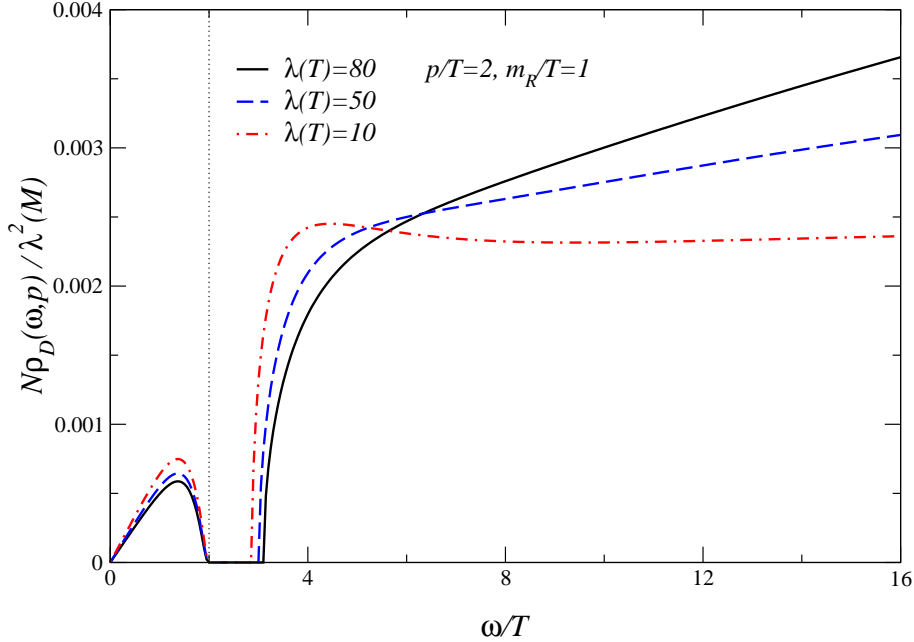}
 \end{center}
 \vspace{-0.5cm}
 \caption{Spectral density $N\rho_D(\om,\pv)/\lambda^2(M)$ as a 
function 
 of $\om/T$ for fixed $p/T=2$, $m_R/T=1$ and three values of the coupling 
 $\lambda(\mu=T)$. The vertical line indicates the lightcone.}
 \label{figDrho}
\end{figure}

The chain of bubbles is summed by the auxiliary correlator $D$ in Eq.\ 
(\ref{eqDeucl}). The divergent piece of the single bubble $\Pi_0(0)$ is 
absorbed by coupling constant renormalization. The renormalized expression 
for $D^{-1}$ reads
\be
D^{-1}(P) =  -3N\left[ \frac{1}{\lambda(\mu)} -\frac{1}{96\pi^2}\ln
\frac{M^2}{\mu^2} \right] + \Pi_0'(P) + \Pi_T(P).
\ee
The auxiliary correlator is renormalization group invariant.
Retarded auxiliary propagators are obtained as $D_R(P)
= D(i\om_n\to p^0+i0,\vecp) = D_A^*(P)$. The spectral density is
\be
\rho_D(P) =  -i\left[D_R(P) - D_A(P) \right] \\
= 
\frac{-2\im\Pi_R(P)}{\left[3N/\lambda(M) - \re\Pi_R^f(P)\right]^2
+ \left[\im\Pi_R(P)\right]^2},
\ee
in terms of the RG invariant coupling
\be
\frac{1}{\lambda(M)} = \frac{1}{\lambda(\mu)} -\frac{1}{96\pi^2}\ln
\frac{M^2}{\mu^2}.
\ee

The spectral function is shown in Fig.\ \ref{figDrho} for a typical
choice of parameters. We scaled out the trivial $N/\lambda^2(M)$
dependence. $\im \Pi_R(\om,\pv)$ and therefore $\rho_D(\om,\pv)$ are 
nonzero below the lightcone ($\om^2<p^2$) and above threshold 
($\om^2>p^2+4M^2$). 
We note that for larger coupling constant the contribution
below the lightcone diminishes. Fixing $m_R$ and increasing $\lambda(T)$ 
results in a larger value for $M$ (see Fig.\ \ref{figmassmR}). As a result 
the contribution above threshold starts at larger $\om$ when 
increasing $\lambda(T)$.


\subsection{Thermal width}  \label{thw}

The thermal width is given by 
\be
 \Gamma_\pv = 
 -\frac{\im \Sigma_R(P)}{p^0}\Big|_{p^0=\pm\om_\pv} = 
 -\frac{\im \Sigma_R(\om_\pv,\pv)}{\om_\pv}.
\ee
The LO part of $\Sigma$ does not contribute. We write the 
self energy at NLO as
\be
 \Sigma^{\rm NLO}(P) = -\si_R G(R)D(R-P).
\ee
\begin{figure}[t]
 \begin{center}
 \epsfig{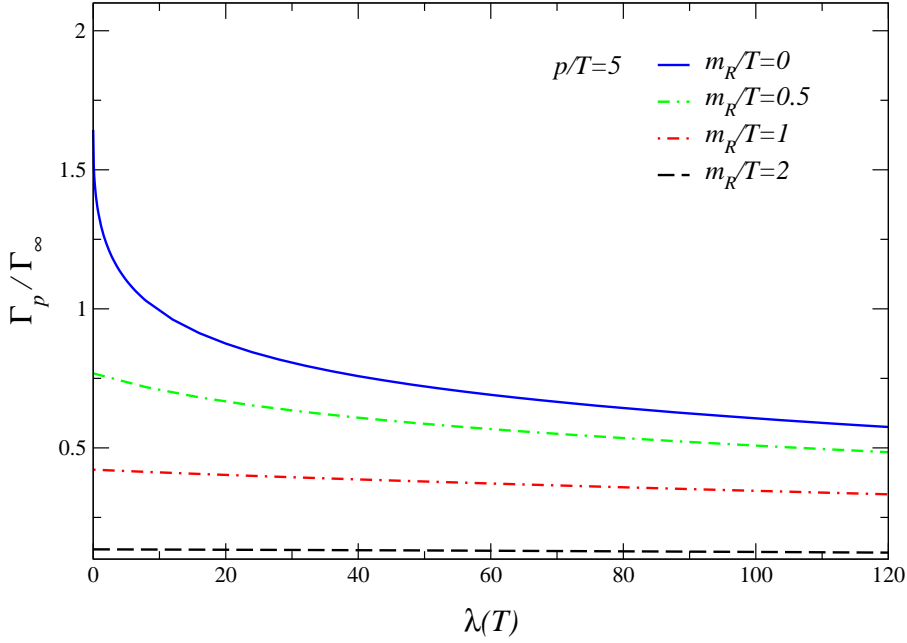}
 \end{center}
 \vspace{-0.5cm}
 \caption{Width $\Gamma_\pv/\Gamma_{\infty}$ as a function 
 of $\lambda(T)$ for fixed $p/T=5$ and four values of $m_R/T$.}
 \label{figwidth2}
\end{figure}
After performing the Matsubara sum and the analytical continuation, we 
find $\im \Sigma_R$ in terms of spectral functions as
\be
 \im \Sigma_R(P) = \frac{1}{2}\int_{R} 
 \rho(R)\rho_D(R-P)\left[n(r^0) - n(r^0-p^0)\right].
\ee
A convenient way to proceed is to introduce $k = |\rv -\pv|$ as 
\be  \label{eqkrp}
 1 = \int_0^\infty dk\,\delta(k-|\rv -\pv|) = \int_{|r-p|}^{r+p} 
 dk\,\frac{k}{rp}\delta(\cos \theta_{pr} -z_{pr}), 
\ee
where $\cos \theta_{pr} = \pvuni\cdot\rvuni$ is the cosine of the angle 
between $\pv$ and $\rv$ and
\be \label{eqzpr}
 z_{pr} = \frac{r^2+p^2-k^2}{2rp}.
\ee
We perform the $r^0$ integral using $\rho(R)$ and the $\theta_{pr}$ 
integral using the delta function introduced above. The final result for 
the width then reads
\bea
\nn
\Gamma_\pv =
\frac{1}{16\pi^2 p\,\om_\pv} 
 \int_0^\infty dr\, \frac{r}{\om_\rv} 
\int_{|r-p|}^{r+p} 
dk\,k 
\Big\{
\rho_D(\om_\rv+\om_\pv,k) 
\left[ n(\om_\rv) - n(\om_\rv+\om_\pv) \right]
&&\hm
 \\
 - \rho_D(\om_\rv-\om_\pv,k) 
\left[ n(\om_\rv) - n(\om_\rv-\om_\pv) \right]
 \Big\}.
&&\hm
\eea
The remaining two integrals can be performed numerically. Note that to 
obtain $\rho_D$, a numerical evaluation of the real part of the 
bubble is required as well. 

In Fig.\ \ref{figwidth2} we show the width as a function of the coupling 
constant for fixed $p/T=5$ and several choices of $m_R/T$.
In order to remove the trivial parameter dependence, we rescaled 
$\Gamma_\pv$ with 
\be
\Gamma_{\infty} = \frac{\lambda^2(M)T^2}{2403\pi N p},
\ee
the thermal width at ultrahard momentum $p\gg T$ in the weakly coupled, 
massless limit at leading order in the $1/N$ expansion (see Sec.~\ref{weakcou}).
We observe that for small mass the dependence on the coupling constant 
is substantial, whereas for larger mass it becomes negligible.


\section{Shear viscosity} \label{eta}

\begin{figure}[b]
 \begin{center}
 \epsfig{figure=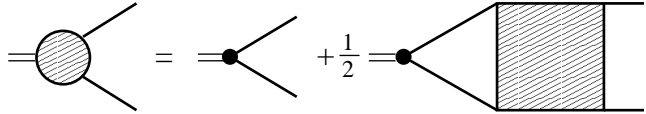,height=1.5cm}
 \end{center}
 \caption{Relation between 3 and 4-point vertex function.}
 \label{figthreefour}
\end{figure}

\begin{figure}[t]
 \begin{center}
 \epsfig{figure=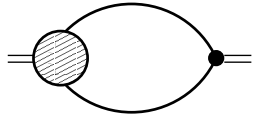,height=1.5cm}
 \end{center}
 \vspace{-0.5cm}
 \caption{The correlator $\langle\pi_{ij}\pi_{ij}\rangle$ in terms of the full vertex.}
 \label{figexpecvertex}
\end{figure}

Now we have all the ingredients necessary to compute the shear viscosity.  
In order to do the Matsubara frequency sums, it is convenient to transform 
the integral equation for the 4-point function into an equivalent one 
for a 3-point vertex, defined in Fig.\ \ref{figthreefour}.
 The corresponding integral equation is depicted in Fig.\ \ref{figvertexN}.
We denote the effective vertex as $\Gamma_{ij,ab}(P+Q,P)$ where $P$ is the
(euclidean) momentum on the siderails and $Q=(i\om_q,\vecnul)$ the
momentum that enters from the left. The integral equation then reads
\be \label{eqintall}
 \Gamma_{ij,ab}(P+Q,P)=\cD_{ij,ab}^{0}(\pv)+
 \frac{1}{2}\si_{R} G(R+Q)\Gamma_{ij,cd}(R+Q,R)G(R)\Lambda_{cd;ab}(R,P;Q),
\ee
where the kernel $\Lambda_{cd;ab}(R,P;Q)$ for vanishing $Q$ was presented 
in Eqs.\ (\ref{eqLambdaLO}, \ref{eqLambdaNLO}) and the bare coupling 
between the scalar field and the $\pi_{ij}$ operator follows from Eq.\ 
(\ref{pi}):
\begin{figure}[t]
 \begin{center}
 \epsfig{figure=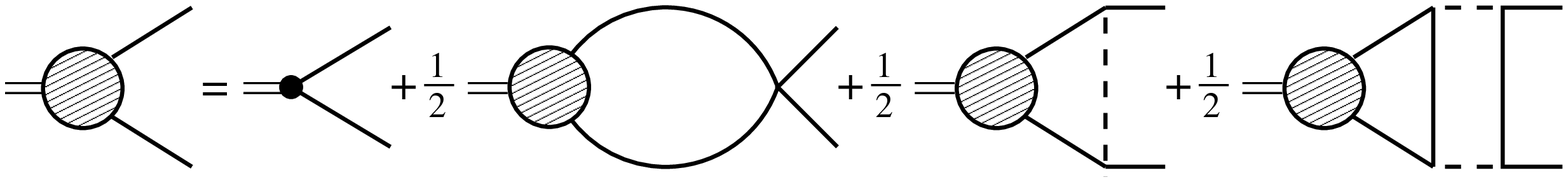,height=1.5cm}
 \end{center}
 \vspace{-0.5cm}
 \caption{Integral equation for the full 3-point vertex.}
 \label{figvertexN}
\end{figure}
\be
\cD_{ij,ab}^0(\pv) = 2\delta_{ab}\left[ p_ip_j - \frac{1}{3}\delta_{ij}
\pv^2 \right].
\ee
The correlator we need to obtain the shear viscosity takes the form of an 
effective one-loop diagram (see Fig.\ \ref{figexpecvertex}),
\be  
 G_{\pi\pi}(Q) = \half\si_{P} 
G(P+Q)\Gamma_{ij,ab}(P+Q,P)G(P)\cD_{ij,ab}^{0}(\pv).
\ee
The effective vertex can be taken of the form\footnote{In principle, 
there could be an additional term proportional to $\delta_{ij}$. However 
this would not contribute when contracted with the
traceless bare vertex.}
\be
 \tw_{ij,ab}(P+Q,P)= 2\delta_{ab}\left[p_{i}p_{j}-\frac{1}{3}\delta_{ij} 
 \pv^{2}\right]\tw(P+Q,P),
\ee
which yields
\be  
\label{eqGpipi}
  G_{\pi\pi}(Q) = \frac{4}{3}N\si_{P} p^{4} G(P+Q)\tw(P+Q,P)G(P).
\ee

In order to do the sums over the Matsubara frequencies, we follow the 
technique presented in Ref.~\cite{ValleBasagoiti:2002ir}, which makes 
use of the following relation
\be
\label{eqvalle}
T\sum_n f(i\om_n) =
\sum_{\rm cuts} \int_{-\infty}^\infty \frac{d\xi}{2\pi i} n(\xi)
{\rm Disc} f - \sum_{\rm poles} n(z_i){\rm Res}(f,z_i),
\ee
which requires knowledge of the analytical structure of the function to be 
summed. Because of the inclusion of the NLO contribution to the self 
energy in the propagators, the propagator $G(\po,\pv)$ has a cut along the 
entire real $p^0$ axis, i.e.\ when $\im(\po)=0$ in the complex $\po$ 
plane. It follows from the integral equation Eq.~(\ref{eqintall}) that the 
effective vertex has the same analytic structure, namely, $\tw(P+Q,P)$ has 
cuts when $\im(\po+\qo)=0$ and $\im(\po)=0$. With this information we can 
do the Matsubara frequency sum in Eq.\ (\ref{eqGpipi}), make the analytical
continuation $i\om_q \to q^0+i0$ afterwards to arrive at the retarded
function and take the limit $q^0\to 0$ (see Refs.\
\cite{ValleBasagoiti:2002ir,Aarts:2002tn} for further illustration). Only
the product of retarded and advanced propagators suffers from pinching
poles and dominates at leading order in the $1/N$ expansion. Therefore
only one particular analytical continuation of the full vertex is needed.
Defining
 \be  \label{anal}
 \cD(p^0,\pv)=\lim_{q^0\to 0}\re\,\tw(p^0+q^0+i0, p^0-i0;\pv),
\ee
the result for the spectral density reads
\be
 \lim_{q_0\to 0} \rho_{\pi\pi}(q^0,\vecnul) = 
 -\frac{8}{3}Nq^0\int_{P} p^{4} n'(p^0) G_R(P) G_A(P) \cD(p^0,\pv).
\ee
Using Eq.\ (\ref{eqPP}) for the product of $G_R$ and $G_A$ and 
definition (\ref{eqetadef}) for the shear viscosity, we get
\be
\eta = -\frac{N}{15}\int_{P} n'(p^0) p^4
\frac{\rho(P)}{2p^0\Gamma_\pv} \cD(p^0,\pv) 
 = -\frac{N}{15}\int_{\pv} n'(\om_\pv) \frac{p^4}{\om_\pv^2}
\frac{\cD(p)}{\Gamma_\pv},
\ee
where we used that $n'(-\om_\pv) = n'(\om_\pv)$ and we defined 
\be
 \label{eqD1}
 \cD(p) =  \cD(\pm\om_\pv,p).
\ee
 To arrive at our final expression for the viscosity, we proceed as in the
case of gauge theories \cite{ValleBasagoiti:2002ir,Aarts:2002tn} and
define the dimensionless quantity
 \be   \label{eqD2}
 \chi(p)=\frac{p^2}{\om_\pv}\frac{\cD(p)}{\Gamma_\pv}.
\ee
The shear viscosity then reads
\be  \label{eqetachi}
 \eta=-\frac{N}{15}\int_{\pv}\frac{p^{2}}{\om_{\pv}}n'(\om_{\pv})\chi(p) 
 =-\frac{N}{30\pi^2}\int_0^\infty dp\,\frac{p^4}{\om_\pv}n'(\om_\pv)\chi(p).
\ee
Since the thermal width is inversely proportional to $N$, it follows 
that $\chi(p) \sim N$ in the large $N$ limit.


\section{Integral equation}  \label{inte}

As we have shown in the previous section, in order to obtain the shear
viscosity, we need a particular analytic continuation of the effective
vertex in the limit $q^0\to 0$ and $p^0=\pm \om_\pv$. In this section we
first explicitly write down the integral equation in this kinematical
limit. We then show how to cast it in a form suitable for a variational
treatment. We also point out the relation between the integral equation 
and kinetic theory.

\begin{figure}[b]
 \begin{center}
 \epsfig{figure=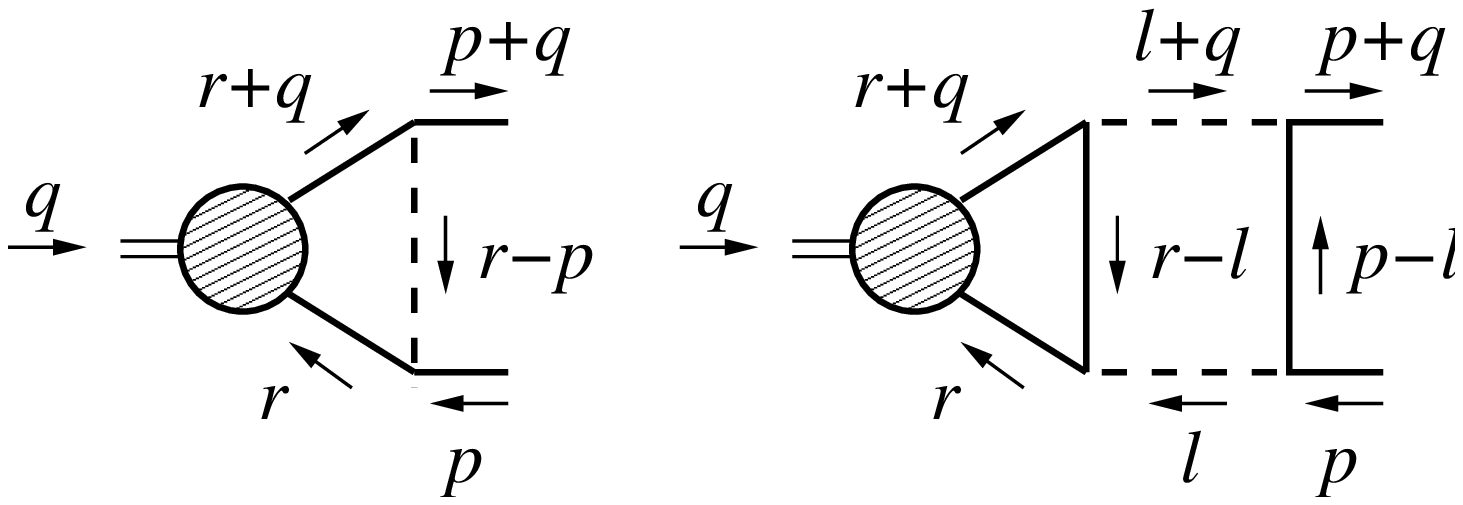,height=3.5cm}
 \end{center}
 \vspace{-0.5cm}
 \caption{Choice of momentum labeling. We also denote $k=|\rv-\pv|$.}
 \label{figmom}
\end{figure}

With momenta flowing as illustrated in Fig.\ \ref{figmom}, the integral
equation reads
\be 
 \Gamma_{ij,ab}(P+Q,P)=\cD_{ij,ab}^{0}(\pv)+
 \frac{1}{2}\si_{R} G(R+Q)\Gamma_{ij,cd}(R+Q,R)G(R)\Lambda_{cd;ab}(R,P;Q),
\ee
where the rung is
\bea
\nn
 \Lambda_{cd,ab}(R,P;Q)=&&\hm-\frac{\lambda_{0}}{3N}\delta_{ab}\delta_{cd}
 +(\delta_{ac}\delta_{bd}+\delta_{ad}\delta_{bc})D(R-P)
 \\
 &&\hm+2\delta_{ab}\delta_{cd}\si_{L}D(L)D(L+Q)G(R-L)G(L-P).
\eea  

To arrive at a scalar equation we contract with $\cD_{ij,ab}^{0}(\pv)$ and 
divide by a common factor to find
\be
 \tw(P+Q,P)=1+\frac{1}{2N}\si_{R}\frac{r^{2}}{p^{2}}P_{2}(\pvuni\cdot\rvuni)
 G(R+Q)\tw(R+Q,R)G(R)\Lambda_{cc,aa}(R,P;Q).
\ee
Here, the second Legendre polynomial $P_{2}(x)=(3x^{2}-1)/2$ arises from 
the contraction
\be
\label{eqcontr2}
\cD_{ij,ab}^0(\pv) \tw_{ij,cd}(R+Q,R) =
\frac{8}{3} p^2r^2P_2(\pvuni\cdot\rvuni)\delta_{ab}\delta_{cd}\tw(R+Q,R),
\ee
 We now perform the Matsubara frequency sums following again the technique
in Ref.~\cite{ValleBasagoiti:2002ir}.  The lowest order term in the rung
does not contribute since it is a constant and it has no discontinuity.
For doing the sum for the second term in the rung, we need to know that
the auxiliary correlator $D(R-P)$ has a branch cut below the lightcone and
above threshold (see Sec.\ 4.2).  However, when using 
Eq.~(\ref{eqvalle}), we may assume that the cut runs 
along the entire real axis, since the contribution from where the cut is 
absent cancels automatically. 
Finally, for the last piece of the rung there are two sums. These can be
performed sequentially, provided that the analytical continuation is made
at the very end. The analytical structure is as follows. The $G$'s that
depend on $R+Q$ and $R$ as well as the vertex have branch cuts along the
entire real energy axis, for the $D$'s we may take branch cuts along the
entire real energy axis, and the $G$'s that depend on $R-L$ and $P-L$
contribute just simple poles (since they do not require the inclusion of 
the thermal width).  When performing the sum over $\om_l$, one picks up 
contributions from the poles of the $G$'s and from the cuts of the $D$'s. 
We found that after making the analytical continuation only the poles from 
the $G$'s contribute. 

{}From Eq.~(\ref{anal}) it follows that we need the following analytic 
continuation
 \be
 i\om_{p} \to p^{0}-i0, \hs{1cm}
 i\om_{q} \to q^{0}+i0, \hs{1cm}
 i\om_{p}+i\om_{q} \to p^{0}+q^{0}+i0.
\ee
Preserving again only the dominant contribution with pinching poles for 
vanishing $\qo$, we arrive at 
\be
 \cD(P)=1+\int_R \frac{r^2}{p^2}P_2(\pvuni\cdot\rvuni)\left[n(r^0-p^0)-n(r^0)\right] 
 G_R(R) \cD(R) G_A(R) \Lambda(R,P),
\ee
with
\be
 \Lambda(R,P) = \rho_D(R-P)+N\int_{L}\left[n(p^0-l^0)-n(r^0-l^0)\right] 
 \rho(P-L) \rho(R-L) \left| D_R(L) \right|^2.
\ee
We note that the kernel obeys 
$\Lambda(R,P)=-\Lambda(P,R)$, $\Lambda(-R,P)=-\Lambda(R,-P)$. 

We now use Eq.\ (\ref{eqPP}) for the product of $G_R$ and $G_A$, Eqs.\ 
(\ref{eqD1}, \ref{eqD2}) to introduce $\chi(p)$, and take 
$p^0=\pm\om_\pv$ (both choices yield the same equation)
to write the integral equation in the final form 
\be
\label{eqfinal}
 \om_\pv \Gamma_\pv \chi(p)  = p^2 + \half\int_R 
 \left[n(r^0-\om_\pv)-n(r^0)\right]  \frac{\om_\rv}{r^0} 
 P_2(\pvuni\cdot\rvuni) \chi(r)\rho(R)\Lambda(R,P).
\ee
Solving for $\chi(p)$, one obtains the shear viscosity from 
Eq.\ (\ref{eqetachi}).


\subsection{Variational approach}

Since the integral equation cannot in general be solved analytically, one
has to rely on numerical methods. A convenient approach to follow is the
one employed in Refs.~\cite{Arnold:2000dr,Arnold:2003zc,Moore:2001fg},
where the problem of obtaining a transport coefficient in kinetic theory
from an integral equation is formulated as a variational problem. Here we
show how to cast the integral equation in a form that is suitable for a
variational treatment.

After multiplying Eq.\ (\ref{eqfinal}) with
\be \label{eqmult}
 \frac{p^2}{\om_\pv}n'(\om_\pv),
\ee
the integral equation can be written rather compactly as
\be \label{eqint}
 {\cal F}(p)\chi(p) = {\cal S}(p)  + \int_0^\infty dr\, {\cal H}(p,r)\chi(r),
\ee
with
\be
 {\cal F}(p) =  p^2 n'(\om_\pv) \Gamma_\pv,
 \;\;\;\;\;\;\;\;\;\;\;\;
 {\cal S}(p) = \frac{p^4}{\om_\pv}n'(\om_\pv), 
\ee
and a symmetric kernel
\be
 {\cal H}(p,r) = {\cal H}(r,p),
\ee
whose explicit form is presented below. Since ${\cal H}$ is symmetric, 
Eq.\ (\ref{eqint}) can be derived by extremizing the functional
\be \label{eqfunc}
 Q[\chi] = \int_0^{\infty} dp\left[
 {\cal S}(p) \chi(p) -\half {\cal F}(p)\chi^2(p)+\half 
 \int_0^\infty dr\, {\cal H}(p,r)\chi(r)\chi(p) \right].
\ee
{}From Eq.\ (\ref{eqetachi}) we see that the actual value of this 
functional 
at the extremum, 
\be
 Q[\chi=\chi_{\rm ext}] = 
 \half \int_0^{\infty} dp\,\,{\cal S}(p)\chi(p),
\ee
is directly proportional to the viscosity
\be
 \eta=-\frac{N}{15\pi^2}Q[\chi=\chi_{\rm ext}].
\ee 
In the rest of this section we explicitly evaluate ${\cal H}(p,r)$. 

We treat separately the diagram with the single line and with the box
diagram and write ${\cal H} = {\cal H}_{\rm line} + {\cal H}_{\rm box}$.
We start with the diagram containing the single line. The integral to
evaluate reads
\be
 \int_R \left[n(r^0-\om_\pv)-n(r^0)\right] \frac{\om_\rv}{r^0} 
 P_2(\pvuni\cdot\rvuni) \chi(r)\rho(R)\rho_D(R-P).
\ee
We proceed in exactly the same way as in the case of the width in 
Sec.~\ref{thw} 
and introduce $k = |\rv -\pv|$ via Eq.\ (\ref{eqkrp}). 
We perform the $r^0$ integral using $\rho(R)$ and the integral over the 
angle between $\pv$ and $\rv$ using the delta function in Eq.\ 
(\ref{eqkrp}). Multiplying the result with Eq.\ (\ref{eqmult}), 
we can immediately read the contribution to ${\cal H}(p,r)$ 
introduced above, and we find
\bea
 {\cal H}_{\rm line}(p,r) =  
\frac{n'(\om_{\pv})}{16\pi^2}\frac{p}{\om_\pv}\frac{r}{\om_{\rv}}
 \int_{|p-r|}^{p+r}dk\, k &&\hm P_{2}(z_{pr})\Big\{
 \rho_D(\om_{\rv}-\om_{\pv},k)
 \left[n(\om_{\rv}-\om_{\pv})-n(\om_{\rv})\right] 
 \nn  \\ &&\hm 
 - \rho_D(\om_{\rv}+\om_{\pv},k)
 \left[n(\om_{\rv}+\om_{\pv})-n(\om_{\rv})\right]\Big\},
\eea
where $z_{pr}$ is given in Eq.\ (\ref{eqzpr}). Using the relations
\be  \label{nrel}
 n(\om_{\rv}+\om_{\pv})-n(\om_{\rv}) 
 = \frac{Tn'(\om_{\rv})}{1+n(\om_{\rv})+n(\om_{\pv})},
 \;\;\;\;\;
 n(\om_{\rv}-\om_{\pv})-n(\om_{\rv}) =
 \frac{Tn'(\om_{\rv})}{n(\om_{\rv})-n(\om_{\pv})},
\ee
and the fact that $\rho_D(\om_\rv-\om_\pv,k)$ is odd under interchange 
of 
$p$ and $r$, it is straightforward to verify that the result is symmetric 
in $p$ and $r$.

For the diagram containing the box rung, we have to evaluate
\bea  \nn
 \int_{R,L} 
 \left[n(r^0-\om_\pv)-n(r^0)\right] \left[n(\om_\pv-l^0)-n(r^0-l^0)\right] 
 && \\ \times
 P_2(\pvuni\cdot\rvuni) \chi(r)\frac{\om_\rv}{r^0}\rho(R)
 \rho(P-L)\rho(R-L)\left| D_R(L)\right|^2. &&
\eea
There are three angular integrations that are nontrivial.
We denote the cosine of the angle between $\pv$ and $\lv$ as 
$\cos \theta_{pl}$, between $\rv$ and $\lv$ as $\cos\theta_{rl}$, and the 
azimuthal angle between the $\pv,\lv$ plane and the 
$\rv,\lv$ plane as $\phi$.
The 8-dimensional integral can then be written as
\be
\frac{2\pi}{(2\pi)^8} 
\int_0^\infty dr\,r^2 \int_{-\infty}^\infty dr^0 
\int_0^\infty dl\,l^2 \int_{-\infty}^\infty dl^0
\int_{-1}^1 d\!\cos \theta_{pl} \int_{-1}^1 d\!\cos\theta_{rl} 
\int_0^{2\pi} 
d\phi.
\ee
The integration over $\cos \theta_{pl}$ will be performed using the
delta functions in $\rho(P-L)$, the one over $\cos \theta_{rl}$ using
$\rho(R-L)$ and that over $r^0$ using $\rho(R)$. The product of the three
spectral functions yields a set of constraints, since
\bea
\nn
\rho(R)\rho(P-L)\rho(R-L)\Big|_{p^0=\om_\pv} 
\!\!\!\!  \sim &&\hm \sum_{s_i=\pm} 
\delta(r^0+s_1\om_\rv) 
\delta(\om_\pv-l^0+s_2\om_{\pv-\lv})
\delta(r^0-l^0-s_3\om_{\rv-\lv}) \\
\sim  &&\hm 
\sum_{s_i=\pm}\delta(\om_\pv+s_1\om_\rv+s_2\om_{\pv-\lv}+s_3\om_{\rv-\lv}).
\eea
Out of the eight combinations, only three can contribute for kinematical 
reasons, namely those corresponding to $2\leftrightarrow 2$ processes. We 
treat these three cases separately and write
\be
{\cal H}_{\rm box} = {\cal H}_{\rm box}^{(1)} +
{\cal H}_{\rm box}^{(2)} + {\cal H}_{\rm box}^{(3)}.
\ee

\begin{enumerate}
\item $(s_1, s_2, s_3) = (-,+, -)$.
The cosines are $\cos \theta_{pl} = z_{pl}$, $\cos \theta_{rl} = 
z_{rl}^-$, where
\be
 z_{pl} = \frac{l^2-l_0^2}{2pl} +\frac{\om_\pv l^0}{pl},
\;\;\;\;\;\;\;\;
 z_{rl}^{s_1} = \frac{l^2-l_0^2}{2rl} -s_1\frac{\om_\rv l^0}{rl}. 
\ee
The constraints from the spectral functions can be satisfied provided 
\be
l^0> \sqrt{l^2+4M^2}, \;\;\;\;\;\;\;\; |l_-| < p,r < |l_+|,
\ee
where here and below
\be
l_\pm = \half \left[l\pm l^0\beta(L)\right], \;\;\;\;\;\;\;\;
\beta(L) = \sqrt{1-\frac{4M^2}{l_0^2-l^2}}.
\ee
The only place where the angle $\phi$ appears is in $\pvuni\cdot\rvuni$, 
which when expressed in terms of the angles we use, reads 
\be
\pvuni\cdot\rvuni =
\sin\theta_{pl}\sin\theta_{rl}\cos\phi
+ \cos\theta_{pl}\cos\theta_{rl}.
\ee
It is then also straightforward to perform the $\phi$-integral and we find
\be
\int_0^{2\pi}d\phi\, P_2(\pvuni\cdot\rvuni) = 2\pi 
P_2(\cos\theta_{pl})P_2(\cos\theta_{rl}).
\ee 

Multiplying the resulting expression with Eq.\ (\ref{eqmult}) we can read 
off the first contribution to ${\cal H}(p,r)$ from the box diagram:
\bea
\nn
&&\hm 
{\cal H}_{\rm box}^{(1)}(p,r) = 
\frac{N}{128\pi^3} \frac{p}{\om_\pv}\frac{r}{\om_\rv} 
n'(\om_\pv)\left[n(\om_\rv-\om_\pv) - n(\om_\rv) \right]
 \\ && \times\nn
\int_0^\infty dl 
\int_{\sqrt{l^2+4M^2}}^\infty dl^0\,
P_2(z_{pl}) P_2(z_{rl}^-) 
\left| D_R(L)\right|^2 \left[n(\om_\pv-l^0)-n(\om_\rv-l^0)\right] 
 \\ && \times
\Theta(p-|l_-|)\Theta(|l_+|-p)\Theta(r-|l_-|)\Theta(|l_+|-r).
\eea

\item $(s_1,s_2,s_3) = (-,-,+)$.
The cosines are $\cos \theta_{pl} = z_{pl}$, $\cos \theta_{rl} = 
z_{rl}^-$ and the constraints are
\be
l_0^2< l^2, \;\;\;\;\;\;\;\; p > |l_+|, \;\;\;\;\;\;\;\; r > |l_+|.
\ee
Therefore the second contribution reads
\bea
&&\hm \nn
{\cal H}_{\rm box}^{(2)}(p,r) = 
\frac{N}{128\pi^3} \frac{p}{\om_\pv}\frac{r}{\om_\rv} 
n'(\om_\pv)\left[n(\om_\rv-\om_\pv) - n(\om_\rv) \right]
 \\ && \times\nn
\int_0^\infty dl 
\int_{-l}^l dl^0\,
P_2(z_{pl}) P_2(z_{rl}^-) 
\left| D_R(L)\right|^2 \left[n(\om_\pv-l^0)-n(\om_\rv-l^0)\right] 
 \\ && \times
\Theta\left(p-|l_+|\right)\Theta\left(r-|l_+|\right).
\eea

\item $(s_1,s_2,s_3) = (+,-,-)$.
In this case the cosines are $\cos \theta_{pl} = z_{pl}$, $\cos 
\theta_{rl} = z_{rl}^+$, with the constraints
\be
l_0^2< l^2, \;\;\;\;\;\;\;\; p > |l_+|, \;\;\;\;\;\;\;\; r > |l_-|.
\ee
The third contribution then reads
\bea
&&\hm \nn
{\cal H}_{\rm box}^{(3)}(p,r) = 
\frac{N}{128\pi^3} \frac{p}{\om_\pv}\frac{r}{\om_\rv} 
n'(\om_\pv)\left[n(\om_\rv+\om_\pv) - n(\om_\rv) \right]
 \\ && \times\nn
\int_0^\infty dl 
\int_{-l}^l dl^0\,
P_2(z_{pl}) P_2(z_{rl}^+) 
\left| D_R(L)\right|^2 \left[n(l^0-\om_\pv)-n(l^0+\om_\rv)\right] 
 \\ && \times
\Theta\left(p-|l_+|\right)\Theta\left(r-|l_-|\right).
\eea

\end{enumerate}
Using relations (\ref{nrel}) and making the substitution $l^0\to -l^0$ in
the third contribution, one finds that ${\cal H}_{\rm box}(p,r)={\cal
H}_{\rm box}(r,p)$. In conclusion, ${\cal H}(p,r) = {\cal H}(r,p)$, which
allows to obtain the integral equation from the functional $Q$.


\subsection{Kinetic theory}

Here we briefly mention the relation between our results and the 
corresponding kinetic theory, by analyzing the kernel in the 
integral equation.

\begin{figure}[t]
 \begin{center}
 \epsfig{figure=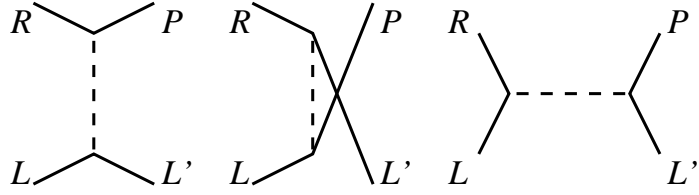,height=2.5cm}
 \end{center}
 \vspace{-0.5cm}
 \caption{Scattering processes in kinetic theory, time runs horizontally.
 }
 \label{figscatt}
\end{figure}

The kernel $\Lambda(R,P)$ can be written in a form that allows for 
a direct comparison with kinetic theory. We start by changing 
variables $L \to P-L$ and find
\be
\Lambda(R,P) = \rho_D(R-P) +
N\int_L \left[n(l^0)-n(r^0-p^0+l^0)\right] \rho(L) \rho(R-P+L) \left| 
D_R(P-L) \right|^2.
\ee
If we use for the spectral density the following identity
\be
\rho_D(R-P) = -2\im \Pi_R(R-P) \left|D_R(R-P)\right|^2,
\ee
where (see Eq.\ (\ref{eqimpi1}))
\be
\im \Pi_R(R-P) =
-\frac{N}{4}\int_{L} \left[ n(l^0) - n(r^0-p^0+l^0) \right]
\rho(L) \rho(R-P+L),
\ee
introduce the variable $L'$ according to
\be
\int_{L'} (2\pi)^4\delta^4(R-P+L-L'),
\ee
and use the delta function to interchange momentum labels, the kernel can 
be written as
\bea
\nn
\Lambda(R,P) = &&\hm \frac{N}{2}\int_{L,L'} \left[n(l^0)-n(l'^0)\right]
\rho(L) \rho(L') (2\pi)^4\delta^4(R-P+L-L')
\\
&&\hm
\times\left[
\left| D_R(R-P) \right|^2 + \left| D_R(R-L') \right|^2 +
\left| D_R(R+L) \right|^2
\right].
\eea
The second line is precisely the sum of the squares of the matrix elements
$|{\cal M}|^2$ for scattering $(R,L) \to (P,L')$ in kinetic theory, where
the auxiliary correlator carries the interaction (see Fig.\
\ref{figscatt}). There are no interference terms (these would appear
from the square of the sum of matrix elements): interference only
contributes at the next order in the $1/N$ expansion.

\section{Weak coupling}  \label{weakcou}

In the limit of weak coupling, the integral equation simplifies
considerably and it is possible to obtain an approximate analytical
solution. At weak coupling, we make the replacements
\be
D_{R/A}(P)\to -\frac{\lambda}{3N}, \;\;\;\;\;\;\;\;
\rho_D(P)\to -\left(\frac{\lambda}{3N}\right)^2 2\im\Pi_R(P),
\ee
so that the kernel becomes
\be
\Lambda(R,P) \to -6\left(\frac{\lambda}{3N}\right)^2 \im\Pi_R(R-P),
\ee
and the width reads
\be
\Gamma_\pv  = \frac{1}{\om_\pv}\left(\frac{\lambda}{3N}\right)^2 
\int_R\rho(R)
\,\im\Pi_R(R-P) \left[ n(r^0) - n(r^0-\om_\pv) \right].
\ee
We will use the integral equation in the form
\be
\om_\pv \Gamma_\pv \chi(p)  = p^2 + \half\int_R 
\left[n(r^0-\om_\pv)-n(r^0)\right]  \frac{\om_\rv}{r^0} 
P_2(\pvuni\cdot\rvuni) \chi(r)
\rho(R)\Lambda(R,P).
\ee
Both on the left and the right-hand side of this equation two integrals
(over $r$ and $k=|\rv-\pv|$) remain to be done. The imaginary part of the
single bubble $\im \Pi_R$ is known analytically, see Eq.\ (\ref{eqimPi}).
For this reason the weakly coupled case is considerably easier than the
full problem.

For the massless weakly coupled theory, we will now show that it is 
possible to find an approximate analytical solution in the limit $p \to 
\infty$. Consider first the thermal width on the LHS. Evaluating 
the two remaining integrals in that limit one finds (see also Appendix G 
in Ref.\ \cite{Jeon:if} for the case $N=1$)
\be
\lim_{p\to \infty} \Gamma_\pv = \frac{1}{2304\pi} 
\frac{\lambda^2T^2}{p N} \left[1+ \frac{24\zeta(3)}{\pi^2}\frac{T}{p}
+ {\cal O}(e^{-p/T})\right].
\ee
Neglecting for a moment the ladder contribution on the RHS of the integral 
equation, it is easy to solve for $\chi(p)$ and we find
\be
\label{eqchino}
\lim_{p\to \infty} \chi(p) = 2304\pi\frac{N}{\lambda^2} \frac{p^2}{T^2} 
\left[ 1 +{\cal O}\left(T/p\right) \right].
\;\;\;\;\;\;\;\;\mbox{(no ladders)}.
\ee
To include the ladders we now use the same momentum dependence for $\chi$
but with an arbitrary prefactor
\be
\chi(p) = \kappa p^2.
\ee
Inserting this ansatz on the RHS of the integral and performing the 
integrals while consistently dropping terms that are suppressed with 
respect to the leading $p^2$ behavior, we find the following result for 
the complete integral equation: 
\be
\frac{\lambda^2T^2}{2304\pi N} \kappa p^2 \left[1+{\cal 
O}(T/p)\right] = p^2 + 
\frac{\lambda^2T^2}{3456\pi N} \kappa p^2 \left[1+{\cal
O}(T/p)\right].
\ee
The solution for ultrahard $p$ is therefore
\be
\label{eqchi0}
\lim_{p\to\infty} \chi(p) =   6912 \pi\frac{N}{\lambda^2}
\frac{p^2}{T^2}.
\ee
A comparison between the asymptotic solution without ladders, Eq.\ 
(\ref{eqchino}), and with ladders,  Eq.\ (\ref{eqchi0}), reveals that the 
effect of summing ladders is simply algebraic: the full vertex ${\cal 
D}(p)$ in Eq.\ (\ref{eqD2}) equals 3.

Using this asymptotic form of $\chi(p)$ in expression (\ref{eqetachi}) 
for the viscosity yields 
\be 
\label{eqetaasymp}
\eta_\infty = \frac{27648\zeta(5)}{\pi}
\frac{N^2T^3}{\lambda^2} \approx 9125.6\frac{N^2T^3}{\lambda^2}, 
\ee 
where the subscript indicates that this solution has been obtained for 
ultrahard momentum $p$.

We can compare this result with the numerical results obtained for the 
shear viscosity in Refs.~\cite{Jeon:if,Arnold:2000dr} for the weakly 
coupled massless $N=1$ theory. In order to do this we must find the 
full $N$ dependence at weak coupling, not just the leading order result 
$\propto N^2$. 
For this we use again the 2PI effective action, but now employing the loop 
expansion to three loops. The important formulas are summarized in 
Appendix A. The result is that the full $N$ dependence of the shear 
viscosity is quite simple, and we find
\be
\eta_\infty = \frac{27648\zeta(5)}{\pi} 
\frac{N^3}{N+2}\frac{T^3}{\lambda^2}
\approx 3041.9  \frac{3N^3}{N+2}\frac{T^3}{\lambda^2}.
\ee
The numerical constant is extremely close to the ones obtained
numerically in Ref.~\cite{Jeon:if} (3040) and
Ref.~\cite{Arnold:2000dr} (3033.5).


\section{Variational solution}  \label{num}

In order to obtain the shear viscosity for general values of the coupling 
constant and mass parameter, we solve the problem of extremizing 
the functional $Q$ in Eq.\ (\ref{eqfunc}) variationally. 
Following Arnold, Moore and Yaffe 
\cite{Arnold:2000dr,Arnold:2003zc,Moore:2001fg},
we expand $\chi(p)$ in a finite set of suitably chosen basis functions 
$\phi^{(m)}(p)$:
\be
 \chi(p) = N \sum_{m=1}^{N_{\rm var}} a_m \phi^{(m)}(p),
\ee
where we factored out an explicit factor of $N$, so that the integrals 
below are $N$-independent. Using this Ansatz in the functional $Q$ yields 
\be
 Q[\chi] = N \sum_m a_m \left[ {\cal S}_m + \half \sum_{n} a_n  
 \left( - {\cal F}_{mn} + {\cal H}_{mn} \right)\right],
\ee
with
\bea \nn
 {\cal S}_m =&&\hm  \int_0^{\infty} dp\,\, 
  {\cal S}(p) \phi^{(m)}(p) , \\
 \label{eqmn}
 {\cal F}_{mn}  =&&\hm N \int_0^{\infty} dp\,\, 
 {\cal F}(p)   \phi^{(m)}(p) \phi^{(n)}(p), \\
 \nn
 {\cal H}_{mn} =&&\hm N \int_0^{\infty} dp\,\int_0^{\infty} dr\, 
 {\cal H}(p,r)  \phi^{(m)}(p) \phi^{(n)}(r).
\eea
 Note that ${\cal S}_m$ is a 1-dimensional integral, ${\cal F}_{mn}$ and
${\cal H}_{mn}$ for the line diagram are 3-dimensional integrals, and
${\cal H}_{mn}$ for the box diagram is a 4-dimensional integral.
Extremizing the functional with respect to the variational parameters
$a_m$ gives a simple linear algebra problem. The shear viscosity is
given by
\be
 \eta=-\frac{N^2}{30\pi^2}\sum_m {\cal S}_m a_m, 
 \hs{1cm}
 a_m = \sum_n ( {\cal F} - {\cal H})^{-1}_{mn} {\cal S}_n.
\ee

\begin{figure}[t]
 \begin{center}
 \epsfig{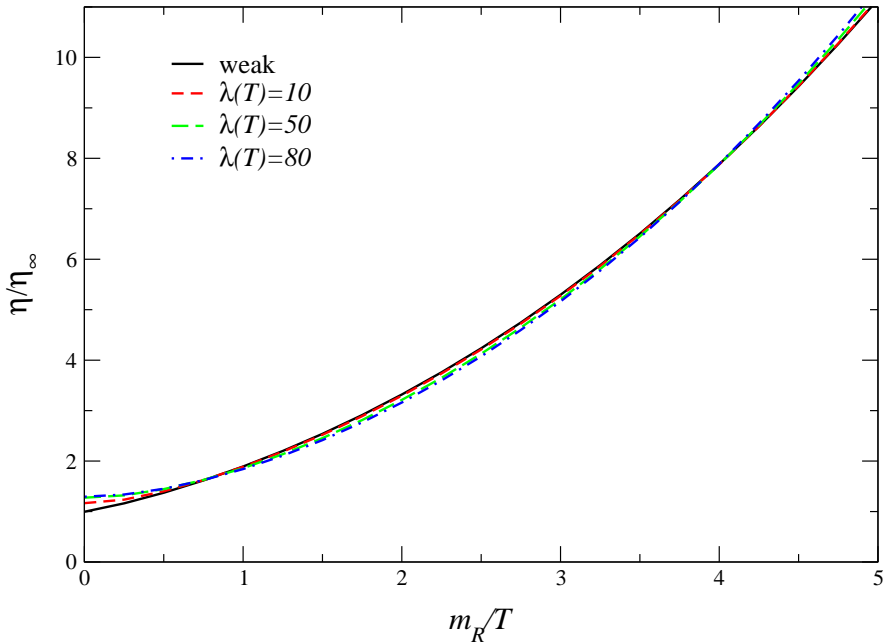}
 \end{center}
 \vspace{-0.5cm}
 \caption{Shear viscosity $\eta/\eta_\infty$ as a function of $m_R/T$, 
 for four values of $\lambda(\mu=T)$. Weak denotes the result in the 
 limit $\lambda\to 0$, i.e. keeping only the single bubble in the rung.} 
 \label{figetamR}
\end{figure}

\begin{figure}[t]
 \begin{center}
 \epsfig{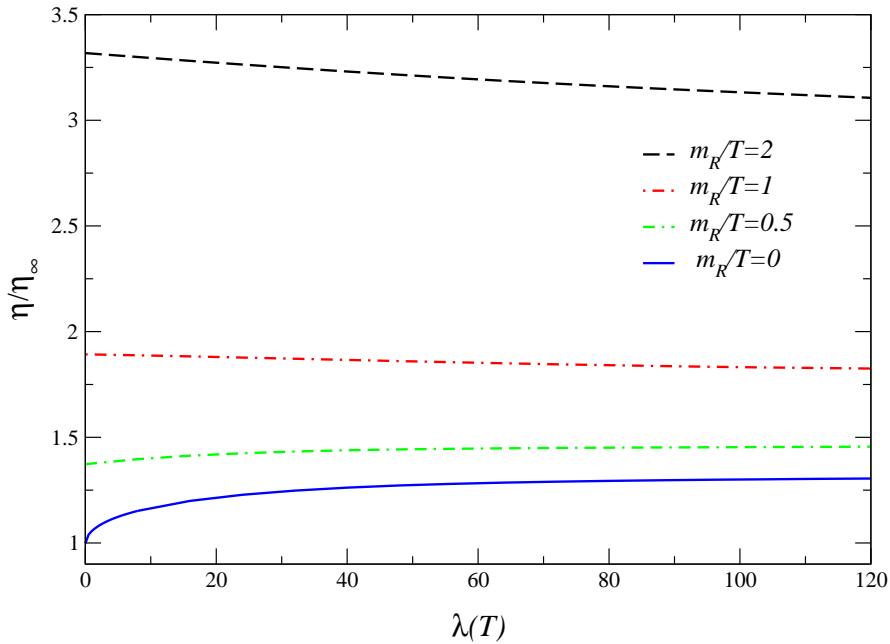}
 \end{center}
 \vspace{-0.5cm}
 \caption{Shear viscosity $\eta/\eta_\infty$ as a function of 
 $\lambda(\mu=T)$, for 4 different values of $m_R/T$.}
 \label{figetalambdaT}
\end{figure}

Given the asymptotic solution $\chi(p) \sim (p/T)^2$, our choice 
for the trial functions is
\be
\phi^{(m)}(p) = \frac{(p/T)^2}{(1+p/T)^{m-1}} 
\sum_{k=0}^{m-1} (-1)^k \left(p/T\right)^k.
\ee
For fixed $N_{\rm var}$ these basis functions are a linear combination of
the functions used in Refs.\
\cite{Arnold:2000dr,Arnold:2003zc,Moore:2001fg}. For given basis functions
the integrals in Eq.\ (\ref{eqmn}) can now be performed numerically. We
found that the presence of the mass $M$ in the integration boundaries for
$p$ and $r$ in ${\cal H}_{\rm box}$ (see Sec.\ 5.2) reduces the effort
required for the numerical integration compared to the massless case.
Using straightforward quadrature, the errors due to the numerical
integration are smaller than the width of the lines in Figs.\
\ref{figetamR}, \ref{figetalambdaT}. Also, the results shown in these
figures are obtained using three basis functions. Again, the effect of 
using this truncated basis set is smaller than the width of the lines.

In Fig.\ \ref{figetamR} we present the shear viscosity as a function of
the renormalized mass in vacuum. To remove the trivial $N^2T^3/\lambda^2$
dependence, we scaled the result with the approximate analytical solution
$\eta_\infty$ (see Eq.\ (\ref{eqetaasymp})).  The line labeled `weak'
shows the result in the weak coupling limit $\lambda\to 0$, i.e.\
preserving only the single bubble in the rung. The other lines represent
the solution to the full problem for various values of $\lambda(T)$.  We
have chosen $\mu=T$ to present our results, but the shear viscosity is RG
invariant. We observe that the shear viscosity has a characteristic
dependence on the mass, but a negligible dependence on the coupling
constant, after the dominant $1/\lambda^2$ behavior has been scaled out.

In order to analyze this in more detail, we show in Fig.\
\ref{figetalambdaT} the shear viscosity as a function of the coupling
constant $\lambda(T)$ for various values of the renormalized mass at zero
temperature. Recall that because of the scaling with $\eta_\infty$, we
concentrate here on the subdominant $\lambda$ dependence. We find an
appreciable dependence on the coupling constant only for vanishing mass.  
Comparison with the thermal width in Fig.\ \ref{figwidth2} shows that for
small $m_R$, the shear viscosity and the inverse width have a similar
dependence on the coupling constant, as expected.


\section{Conclusions}

We have presented a diagrammatic calculation of the shear viscosity in the
$O(N)$ model at first nontrivial order in the large $N$ limit. The $1/N$
expansion of the 2PI effective action at next-to-leading order leads in a
straightforward manner to the diagrams contributing to the shear viscosity
at leading order and provides automatically the integral equation required
to sum them.

In the weakly coupled massless theory, the integral equation could be 
analyzed analytically in the limit of ultrahard momentum. Using this 
result we found for the shear viscosity in the $O(N)$ model
\be
\eta_\infty = \frac{9216\zeta(5)}{\pi} 
\frac{3N^3}{N+2}\frac{T^3}{\lambda^2} \approx 3041.9 
\frac{3N^3}{N+2}\frac{T^3}{\lambda^2}.
\ee
This result is extremely close to the numerical values determined earlier 
for the single component theory ($N=1$). 

For the general case we computed the shear viscosity numerically through a
variational approach. The results are presented in Figs.~\ref{figetamR}
and \ref{figetalambdaT}. Factoring out the basic $1/\lambda^{2}$
dependence, the results show that the remaining dependence on the coupling
constant is very weak, while the effect on the mass parameter is larger.
For the allowed range of parameters, we conclude that the shear viscosity 
is close to the result obtained in the weak-coupling analysis.

{}From a more general point of view, we found that the availability of an
effective action that sums the appropriate diagrams is extremely useful.  
While the computation of transport coefficients is still quite involved,
it is streamlined by the organization inherent in the 2PI effective
action. From the wider perspective of nonequilibrium quantum fields, it is
satisfying that 2PI effective action techniques can be applied both far
from equilibrium, relying mostly on numerical tools, as well as (very)  
close to equilibrium, as we have demonstrated here.

As we have argued recently \cite{Aarts:2003bk}, the first nontrivial
truncation of the 2PI effective action also sums the relevant diagrams to
obtain the shear viscosity and the electrical conductivity in gauge
theories to leading-logarithmic order in the weak coupling limit or to
leading order in a $1/N_f$ expansion, where $N_f$ is the number of fermion
fields. A diagrammatic analysis of transport coefficients in gauge
theories beyond the leading-log approximation at weak coupling is not yet
available. A convenient starting point for such an analysis may be based
on a more general $n$PI effective action approach~\cite{Berges:2004pu}.


\vspace*{0.5cm}
\noindent
{\bf Acknowledgments.} Discussions with U.\ Heinz are gratefully 
acknowledged. This research was supported in part by the 
U.~S.\ Department of Energy under Contract No.\ DE-FG02-01ER41190 and  
No.\ DE-FG02-91-ER4069 and by the Basque Government and in part by 
the Spanish Science Ministry (Grant FPA 2002-02037) and the University 
of the Basque Country (Grant UPV00172.310-14497/2002).


\appendix 


\section{Three-loop expansion}

In order to find the full $N$ dependence of the shear viscosity in the 
$O(N)$ model in the weak coupling limit and not just its leading behavior 
at large $N$, we summarize here the results for the three-loop expansion 
of the 2PI effective action. We write $\Gamma_2[G] = 
\sum_{l=2}^\infty\Gamma_2^{(l)}[G]$, with
\bea
\Gamma_2^{(2)}[G] = &&\hm -\frac{\lambda}{8}\frac{N+2}{3} \int_x 
G^2(x,x),\\
\Gamma_2^{(3)}[G] = &&\hm \frac{i\lambda^2}{48}\frac{N+2}{3N} 
\int_{xy} G^4(x,y).
\eea
 Since mass and coupling constant renormalization do not enter here, we
denote the coupling constant with $\lambda$. We also immediately
specialized to $G_{ab}=\delta_{ab}G$. The corresponding self energies are
\bea
\Sigma^{(2)}_{ab}(x,y) =&&\hm 
- i\frac{\lambda}{2} \frac{N+2}{3N}  \delta_{ab} G(x,x) \delta_{\cal 
C}(x-y),\\
\Sigma^{(3)}_{ab}(x,y) =&&\hm 
- \frac{\lambda^2}{6} \frac{N+2}{3N^2} \delta_{ab} G^3(x,y),
\eea
and the kernel reads
\bea
\nn
\Lambda^{(2)}_{ab;cd}(x,y;x',y') = &&\hm -i\frac{\lambda}{3N}
\left[ \delta_{ab}\delta_{cd} + \delta_{ac}\delta_{bd} + 
\delta_{ad}\delta_{bc} \right] \\
&&\hm \times
\delta_{\cal C}(x-y)\delta_{\cal C}(x-y')\delta_{\cal C}(x'-y),\\
\nn
\Lambda^{(3)}_{ab;cd}(x,y;x',y') = &&\hm -
\frac{\lambda^2}{18N^2} 
\left[ 4\delta_{ab}\delta_{cd} + (N+6) ( \delta_{ac}\delta_{bd} +
\delta_{ad}\delta_{bc}) \right] \\
&&\hm\times 
G^2(x,y) \delta_{\cal C}(x-x')\delta_{\cal C}(y-y').
\eea
In particular we find that
\be
\Lambda^{(3)}_{aa;cc}(x,y;x',y') =  - \lambda^2\frac{N+2}{3N}
G^2(x,y) \delta_{\cal C}(x-x')\delta_{\cal C}(y-y').
\ee
Using these expressions it is straightforward to find the $N$ dependence
of the shear viscosity in the $O(N)$ model for arbitrary $N$. Starting
from the large $N$ result in the weak coupling limit, the thermal width 
$\Gamma_\pv$ and the kernel $\Lambda$ are related to the thermal width and 
the kernel presented in Sec.\ \ref{weakcou} as
\bea
\Gamma_\pv \bigg|_{{\rm arbitrary}\, N} =&&\hm
 \frac{N+2}{N} \Gamma_\pv \bigg|_{{\rm large}\, N}, \\
\Lambda(R,P) \bigg|_{{\rm arbitrary}\, N} =&&\hm
 \frac{N+2}{N} \Lambda(R,P) \bigg|_{{\rm large}\, N}.
\eea
The $N$ dependence that subsequently appears in the integral equation
can be absorbed in $\chi(p)$ as
\be 
\chi(p) \bigg|_{{\rm arbitrary}\, N} =
 \frac{N}{N+2} \chi(p) \bigg|_{{\rm large}\, N}, \\
\ee
so the final effect in the viscosity is quite simple:
\be
\eta \bigg|_{{\rm arbitrary}\, N} =
 \frac{N}{N+2} \eta \bigg|_{{\rm large}\, N}. \\
\ee
In the weakly coupled massless limit we arrive therefore at the  
analytical result,
\be
\eta_\infty = \frac{9216\zeta(5)}{\pi} 
\frac{3N^3}{N+2}\frac{T^3}{\lambda^2},
\ee
for the shear viscosity in  the $O(N)$ model.


\end{document}